\documentclass[9pt,twocolumn,twoside]{osajnl}
\journal{pr} 
\setboolean{shortarticle}{false}
\doi{}

\usepackage{array}
\newcommand{\ii}{{\rm i}}
\newcommand{\RE}{{\rm Re}}
\newcommand{\IMG}{{\rm Im}}
\newcommand{\snc}{\text{sinc}}

\usepackage{braket}
\usepackage{amsmath}
\usepackage{mathtools}
\usepackage{comment}
\usepackage{physics}
\usepackage{dsfont}

\usepackage{graphicx}
\usepackage{hyperref}
\hypersetup{pdfstartview={FitH},pdfpagemode={UseNone},
            colorlinks,linkcolor=blue, citecolor=blue, urlcolor=blue,
            bookmarksopen=true, pdfnewwindow=true}
\newcommand{\email}[1]{\href{maito:#1}{#1}}
\usepackage[all]{hypcap}

\graphicspath{{pict/}{figures/}{}}

\usepackage[english]{babel}

\title{Nonlinear quantum spectroscopy with Parity-Time symmetric integrated circuits}
\author[1,2,*]{Pawan Kumar}
\author[1]{Sina Saravi}
\author[1,2]{Thomas Pertsch}
\author[1]{Frank Setzpfandt}
\author[3,4]{Andrey~A.~Sukhorukov}
\affil[1]{Institute of Applied Physics, Abbe Center of Photonics, Friedrich Schiller University Jena, Albert-Einstein-Str. 15, 07745 Jena, Germany}
\affil[2]{Fraunhofer Institute for Applied Optics and Precision Engineering, Albert-Einstein-Str. 7, 07745 Jena, Germany}
\affil[3]{Research School of Physics, Australian National University, Canberra, ACT 2601, Australia}
\affil[4]{ARC Centre of Excellence for Transformative Meta-Optical Systems (TMOS),
Australia}
\affil[*]{Corresponding author: \email{pawan.kumar@uni-jena.de}}

\begin{abstract}
We propose a novel quantum nonlinear interferometer design that incorporates a passive PT symmetric coupler sandwiched between two nonlinear sections where signal-idler photon pairs are generated. The PT-symmetry enables efficient coupling of the longer-wavelength idler photons and facilitates the sensing of losses in the second waveguide exposed to analyte under investigation, whose absorption can be inferred by measuring only the signal intensity at a shorter wavelength where efficient detectors are readily available. 
Remarkably, we identify a new phenomenon of sharp signal intensity fringe shift at critical idler loss values, which is distinct from the previously studied PT-symmetry breaking. We discuss how such unconventional properties arising from quantum interference can provide a route to enhancing the sensing of analytes and facilitate broadband spectroscopy applications in integrated photonic platforms.
\end{abstract}
\setboolean{displaycopyright}{true}

\begin{document}
\maketitle
\section{Introduction}
The generation of photon pairs through spontaneous parametric down-conversion (SPDC) of a pump photon into signal (s) and idler (i) photons inside a quadratic nonlinear medium is inherently probabilistic and characterized by a quantum mechanical pair-generation probability amplitude \cite{Klyshko:1988:PhotonsNonlinear}. When two such sources of photon pairs constitute a quantum optical system, the resulting signal (and idler) photon amplitude after superposition from the sources does not, in general, show any first-order interference. However, if these sources are pumped by a common coherent pump laser and their idler modes are properly aligned, so that it is impossible to ascertain in which source the photon pair was created, the final signal photon intensity after superposition does show interference \cite{Wang:1991-4614:PRA, Zou:1991-318:PRL}. Indistinguishability of the two sources with regard to pair generation lies at the heart of this quantum interference effect \cite{Zou:1991-318:PRL, Wiseman:2000-245:PLA, Lahiri:2017-33816:PRA}. The idler mode from the first nonlinear source must pass through the second source to ensure this indistinguishability and "induce" the coherence between their signal modes necessary for first-order interference. Such a configuration of two aligned nonlinear sources of photon pairs is commonly referred to as a quantum nonlinear interferometer \cite{Chekhova:2016-104:ADOP, Ou:2020-80902:APLP, Caves:2020-1900138:ADQ, Ferreri:2021-461:QUA}.


Recently, quantum nonlinear interferometers have been employed to perform spectroscopic and imaging applications in technologically challenging spectral ranges like infrared and terahertz \cite{Kalashnikov:2016-98:NPHOT, Lemos:2014-409:NAT, Paterova:2018-25008:QST, Paterova:2020-82:LSA, Paterova:2017-42608:SRP, Valles:2018-23824:PRA, Lindner:2020-4426:OE, Kutas:2020-eaaz8065:SCA}. The exciting aspect of these applications is that detection is only required in the visible spectral range on the shorter wavelength signal photon of correlated pairs emitted from non-degenerate SPDC sources. Most applications thus far have employed bulk optical platforms with nonlinear crystals as the source of photon pairs. An alternative approach is to use photonic integrated circuits to realize on-chip nonlinear interferometers, where nonlinear waveguides serve as the source of photon pairs \cite{Ravaro:2008-151111:APL, Solntsev:2018-21301:APLP, Kumar:2020-53860:PRA, Ono:2019-1277:OL}. This allows to combine the inherent advantages of integrated platforms, such as higher nonlinear conversion efficiency, smaller device footprint, long-term stability, and scalability, with the application prospects of quantum nonlinear interferometry. 

On the other hand, integrated systems implementing advanced physical concepts can also enhance the capabilities of optical interferometers. One particularly promising concept explored in integrated optical devices is parity-time (PT) symmetry~\cite{Guo:2009-93902:PRL, Ruter:2010-192:NPHYS}. In the context of coupled waveguide systems, such as directional couplers and coupled resonators, PT symmetry is usually realized by judiciously incorporating balanced optical losses and/or gain in distinct parts of the system in association with a symmetric refractive index distribution \cite{El-Ganainy:2007-2632:OL, Makris:2008-103904:PRL, El-Ganainy:2018-11:NPHYS}. The intriguing aspect of such a PT symmetric system is that it is characterized by a phase transition phenomenon symbolized by the existence of a PT symmetry breaking loss strength. This means that its response below and above this exceptional point is qualitatively different and forms the basis for a slew of novel effects \cite{Guo:2009-93902:PRL, Ruter:2010-192:NPHYS, Lin:2011-213901:PRL, Hodaei:2014-975:SCI}. Importantly, in recent works PT symmetry has been put to use to enhance the sensitivity of optical structures to external perturbations \cite{Wiersig:2014-203901:PRL, Hodaei:2017-187:NAT}, paving the way for integrated spectroscopic and sensing applications with increased responsivity.

\begin{figure}
\centering
\includegraphics[width = \linewidth]{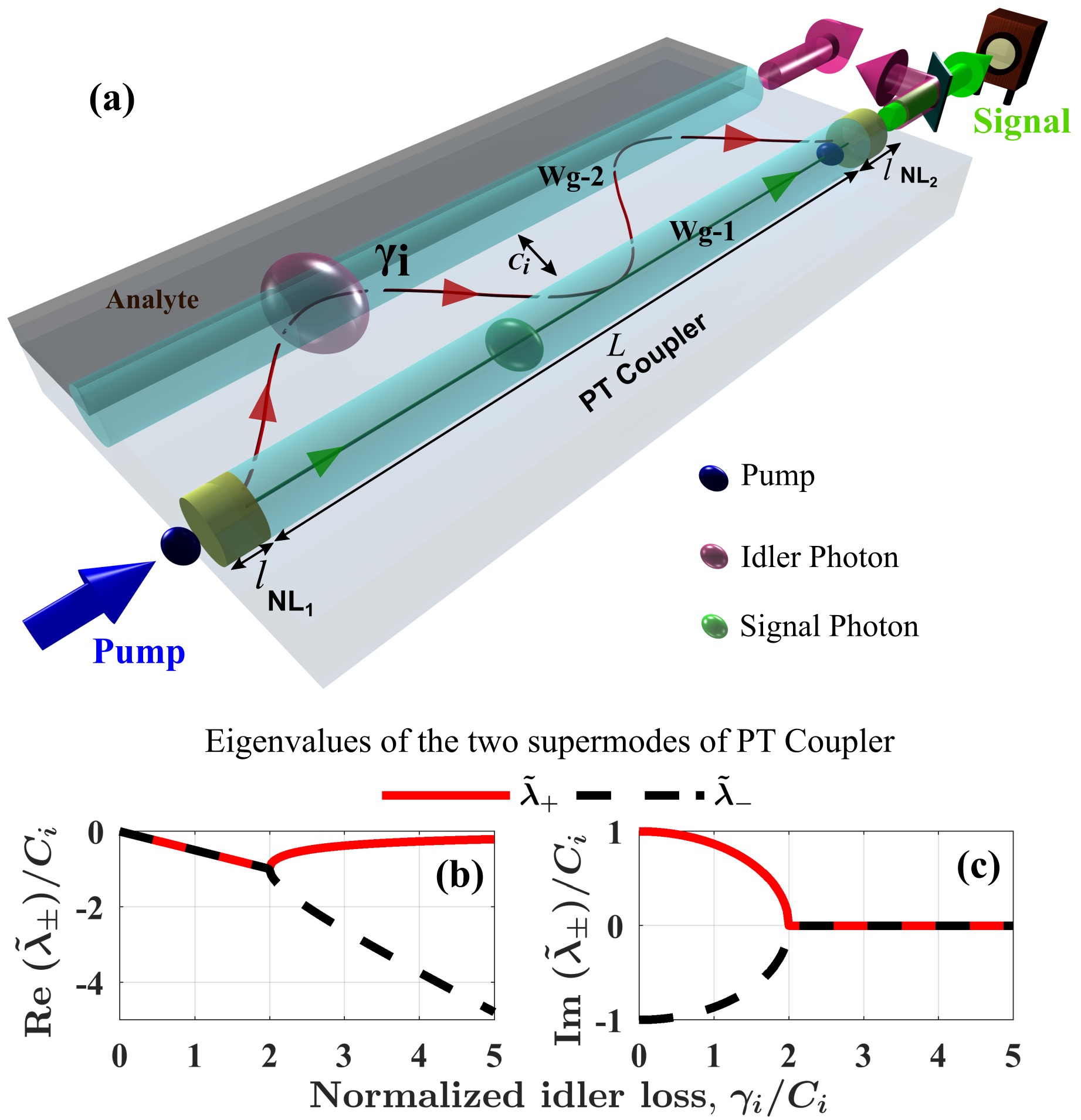}
\caption{(a) Sketch of hybrid nonlinear interferometer incorporating a PT coupler for sensing of analyte-induced absorption ($\gamma_i$) in waveguide Wg-2.
(b, c) Dependence of (b) real and (c) imaginary parts of normalized eigenvalues,  $ \Tilde{\lambda}_\pm/C_i$,
which account for the loss and wavenumber respectively, on the normalized loss $\gamma_i/C_i$ for the two idler supermodes of symmetric PT coupler composed of Wg-1 and Wg-2.}
\label{schematic}
\end{figure}

Here, we propose 
a conceptually new quantum nonlinear interferometer for spectroscopic applications using an integrated waveguide platform with PT symmetry. Different to the standard spectroscopic implementations, the analyte to be probed forms part of a passive PT symmetric coupler \cite{Guo:2009-93902:PRL, Ornigotti:2014-65501:JOPT}. This coupler is positioned between two sources of photon pairs constituting the interferometer [see Fig.~\ref{schematic}(a)]. The coupler offers exceptional control over linear light dynamics such that the output signal intensity can exhibit strong dependence on the idler loss introduced by the analyte in the second waveguide. We show that this opens up new possibilities to tailor the response of the nonlinear interferometer. We identify a new phenomenon of sharply shifting signal interference fringes at specific critical loss magnitudes, which is unrelated to the PT-symmetry breaking effect.
This feature appears due to modulation of quantum indistinguishability between the sources of photon pairs, and can provide a new route to enhancing the sensing of analytes.
The scheme outlined here could also be beneficial in efforts to engineer biphoton quantum states using integrated waveguides and optical-fiber-based nonlinear interferometers \cite{Main:2019-53815:PRA, Su:2019-20479:OE, Ono:2019-1277:OL, Li:2020-204002:APL}.

\section{Nonlinear interferometer integrating a PT-symmetric coupler}
\subsection{Design of interferometer}
We consider an integrated nonlinear interferometer based on waveguides (Wg) as shown in Fig.~\ref{schematic}(a). It consists of a pair of evanescently coupled Wgs where Wg-2 interacts with an analyte that can introduce losses to form a passive PT coupler.
Two short sections of Wg-1, shown in yellow and denoted by NL$_1$ and NL$_2$, possess a quadratic nonlinear susceptibility and function as SPDC sources of signal-idler photon pairs when pumped by a continuous wave laser; here shown as the Pump. On the other hand, the section of Wg-1 between NL$_1$ and NL$_2$ acts as a purely linear element. In practice, such a configuration of linear and nonlinear elements can be easily realised within the same physical Wg made out of quadratic nonlinear materials by engineering the spatial distribution of their effective nonlinearity through quasi-phase matching (QPM) \cite{Hum:2007-180:CRP}. For instance, sections NL$_1$ and NL$_2$  could be periodically poled to modulate their nonlinearity at period $ \Lambda $, while the portion between these could be left unpoled. Assisted by the additional grating vector, $ K = 2\pi/\Lambda $, arising from the QPM poling, the SPDC process can occur efficiently in the two poled sections while the central section merely acts as a dispersive linear element. 

Whereas a simple device geometry employing two spatially separated sources of photon pairs in a single Wg (here Wg-1) can already function as a nonlinear interferometer \cite{Kumar:2020-53860:PRA}, the key suggestion of this work is to introduce a second linear waveguide (Wg-2) adjacent to Wg-1. We show that this enriches the response of the nonlinear interferometer by making use of the exceptional dispersive properties of the resulting PT coupler.

Throughout this paper, we consider the operation of SPDC in the nonlinear sections in the non-degenerate wavelength regime such that the idler photon wavelength is $\lambda_i > \text{2 } \mu m$ while the pump and signal wavelengths lie in the $\lambda_{p,s} < \text{1 } \mu m$ range. Although the exact wavelengths of operation can be suitably chosen as needed by the application, the ranges we indicate here are meant to convey the intended utility of the scheme in facilitating spectroscopic applications in the technologically important mid-infrared spectral range with detection and excitation wavelengths being in the  visible or near-infrared range. 

A salient feature of our proposed design is that the separation distance between Wg-1 and Wg-2 can be suitably chosen so that the longer wavelength idler photon generated in NL$_1$ can tunnel to Wg-2 with coupling rate $C_i$ and evanescently interact with the analyte deposited in its vicinity. At the same time, the shorter wavelength pump and signal photons remain confined to Wg-1 and do not interact with the analyte. However, the idler photon can tunnel back into Wg-1 ensuring that the bi-photon amplitude emanating from NL$_1$ can interfere with that from NL$_2$. Due to the nonlocal nature of this quantum interference, the effective absorption strength in Wg-2, denoted as $\gamma_i$, can be determined by counting just the signal photons at the output of Wg-1. Note that neither the pump nor the signal photons ever propagate through Wg-2 in this scheme.

\subsection{PT Coupler} \label{SECT2:PT_copler}
Before we discuss the generation and evolution of photon pairs in the complete interferometer structure shown in Fig.~\ref{schematic}(a), we provide a brief description of the evolution of classical optical fields within the linear PT coupler between the photon-pair sources. Since the pump and signal fields remain confined to Wg-1, the evolution of their field amplitudes, $a_{p,s}(z)$, is described by $ {\partial a_{p,s}(z)}/{\partial z}  = -\ii \beta_{p,s} \cdot a_{p,s}(z)$ with $\beta_{p,s}$ being their respective propagation constants. On the other hand, the idler field can couple to Wg-2 and, therefore, its field amplitude is described by a two-dimensional vector, $ \mathbf{A}_i(z) = \begin{bmatrix}
a_i^{(1)}(z) & a_i^{(2)}(z)
\end{bmatrix}^\text{T}$ whose evolution is given by
\begin{equation} \label{PT_evol_matrix}
\frac{\partial \mathbf{A}_i(z)}{\partial z}  = \mathbf{M}_i \mathbf{A}_i(z),
\end{equation}
with
\begin{equation}
\mathbf{M}_i  =
\begin{bmatrix}
 - \ii(\Bar{\beta}_i + \Delta \beta_i) & -\ii C_i \\
-\ii C_i &   -\ii(\Bar{\beta}_i - \Delta \beta_i) - \gamma_i
\end{bmatrix}.
\end{equation}
Here, $\Bar{\beta}_i \equiv \left( \beta^{(1)}_i + \beta^{(2)}_i \right)/2$ and $\Delta \beta_i  \equiv \left( \beta^{(1)}_i - \beta^{(2)}_i \right)/2$ are defined in terms of the idler propagation constants $\beta^{(1,\ 2)}_i$ of the two uncoupled Wgs. The loss coefficient $\gamma_i$ is responsible for idler loss in the coupler and is present exclusively in Wg-2. Finally,  $C_i$ denotes the coupling constant between Wg-1 and Wg-2. Solutions of Eq.~(\ref{PT_evol_matrix}) can be obtained analytically by making use of the eigenvalues $ \lambda_\pm = -\ii \Bar{\beta}_i  + \Tilde{\lambda}_\pm$ and corresponding eigenvectors $\mathbf{V_\pm}$ of the coupler propagation matrix $\mathbf{M}_i$, which are
\begin{equation} \label{eigen_val_vect}
\begin{split}
\Tilde{\lambda}_\pm = &\frac{(\pm \xi_i - \gamma_i)}{2}, \\
 \mathbf{V_+} =
&\begin{bmatrix}
\gamma_i +  \xi_i - 2\ii  \Delta \beta_i   &&   -2\ii C_i
\end{bmatrix}^\text{T}, \\
\mathbf{V_-} = &\begin{bmatrix}
 2\ii C_i  &&  \gamma_i +  \xi_i - 2\ii  \Delta \beta_i
\end{bmatrix}^\text{T},
\end{split}
\end{equation}
where $\xi_i  = \sqrt{(\gamma_i - 2\ii \Delta \beta_i)^2 - (2C_i)^2}$ and
 \begin{equation} \label{Formal_sol}
    \mathbf{A}_i(z) = \alpha_+e^{(\lambda_+z)}\mathbf{V_+}  +    \alpha_-e^{(\lambda_-z)}\mathbf{V_-}.
\end{equation}
The coefficients $ \alpha_\pm $ are determined by the input fields to the coupler.
The eigenvectors $ \mathbf{V_\pm} $ describe the two supermodes of the coupler in terms of the individual modes of the two Wgs. The corresponding eigenvalues define their complex propagation constants, which depend on the idler loss magnitude $\gamma_i$ in a nontrivial manner as is evident from the expression for $\Tilde{\lambda}_\pm $. 

We display the characteristic dependence of $\RE\left(\Tilde{\lambda}_\pm \right)$ and $\IMG\left(\Tilde{\lambda}_\pm \right)$, which describe the loss and wavenumber of the supermodes for a symmetric coupler ($ \Delta \beta_i = 0 $), in Figs.~\ref{schematic}(b, c). 
The effect of PT symmetry breaking~\cite{Guo:2009-93902:PRL} occurs at $\gamma_i / C_i = 2$. For idler losses below this threshold value, the two supermodes of the PT coupler have the same effective losses equal to $\gamma_i/2$ (see Fig.~\ref{schematic}(b)). This symmetry between supermodes $\mathbf{V_\pm}$ is due to the nature of their composition in terms of individual modes of the two Wgs below the threshold loss, with $ \norm{\mathrm{V}^{(1)}_\pm} = \norm{\mathrm{V}^{(2)}_\pm}$.  Above the PT symmetry breaking point, $\mathbf{V_+}$ starts to localize in Wg-1 with $\norm{\mathrm{V}^{(1)}_+} > \norm{\mathrm{V}^{(2)}_+}$ while $\mathbf{V_-}$ localizes in Wg-2. Consequently, the loss coefficient for the $\mathbf{V_+}$ supermode decreases with increasing $\gamma_i$, while the opposite is true for the $\mathbf{V_-}$ supermode. As we demonstrate in the following, this has a characteristically defining effect on the behaviour of the proposed nonlinear interferometer. Another important feature to note here is the dependence of $\mathrm{Im}(  \Tilde{\lambda}_\pm ) $ on $\gamma_i$ which is shown in Fig.~\ref{schematic}(c). Both supermodes of the coupler have loss dependent wavenumbers only below the PT symmetry breaking point. In the broken symmetry regime, they become independent of $\gamma_i$.

\section{Methods and Results} \label{sec:methods}
After reviewing the linear properties of the PT-coupler, we now show its profound influence on photon-pair generation in a nonlinear interferometer and discuss applications of the ensuing interference phenomena for quantum sensing. To mathematically describe the behavior of the proposed nonlinear interferometer, we draw upon the theoretical formulation developed in Refs. \cite{Antonosyan:2014-43845:PRA, Antonosyan:2018-A6:PRJ} for describing the process of photon pair generation through SPDC and its propagation in lossy media. We simplify the analysis of pair generation in the nonlinear sections NL$_1$ and NL$_2$ by assuming that their lengths $l$ are much smaller than the coupling length for the idler photon ($l \ll \pi/2C_i$). Hence, in the following we neglect the effect of coupling between Wg-1 and Wg-2 when describing the evolution of signal-idler biphoton amplitudes in the nonlinear sections.
\subsection{Biphoton state amplitude evolution}
We calculate biphoton amplitudes for the quantum state in the Wg-number basis. This is primarily because this basis provides a clear understanding of the spatial dynamics of the generation and propagation of the photon pairs. The  state is described by three probability amplitudes \cite{Antonosyan:2014-43845:PRA} corresponding to the three possible localizations of the signal and idler photons:
\begin{align*}
    &\phi_{11}: \text{signal-idler pair is jointly present in Wg-1,} \\
    &\phi_{12}: \text{signal is present in  Wg-1 while idler is  in Wg-2,} \\
    &\phi_{12}^{\text{(s)}}: \text{signal is in  Wg-1 and idler photon is absorbed  in Wg-2}.
\end{align*}
In this section, we first focus on the evolution of the pair amplitudes $\boldsymbol{\phi} = \begin{bmatrix}
\phi_{11} & \phi_{12}
\end{bmatrix}^\text{T}$, the evolution of $ \phi_{12}^{\text{(s)}} $ is discussed in the subsequent section.
The Schrödinger-type equations for evolution of $\phi_{11}(z)$ and $\phi_{12}(z)$ in each of the nonlinear sections are given by \cite{Belsley:2020-28792:OE}
\begin{equation}\label{NL_sect_EQ}
  \frac{\partial \phi_{1,n_i}(z)}{\partial z}  = -\ii \left(\beta_s + \beta^{(n_i)}_i  \right)\phi_{ 1,n_i}(z) + \delta_{1,n_i} A_p(z),
\end{equation}
where the subscript $n_{i}$ refers to the number of the Wg, 1 or 2, in which the idler photon could be present while its partner signal photon is confined to Wg-1, and the Kronecker delta function $\delta_{1,n_i}$ represents the fact that pair generation can occur only in Wg-1 where the classical pump field given by $A_{p}(z) = A_p(0)\times\cos(K z)\times \exp(-\ii \beta_p z )$ is present. Here, $A_p(0)$ subsumes the strength of the quadratic nonlinearity and the factor $ \cos(Kz) $ represents the dominant spatial modulation of the nonlinearity that realizes QPM in the two nonlinear sections with $K = 2\pi/\Lambda$. Poling period $\Lambda$ is chosen such that $\overline{\beta}_p - \overline{\beta}_s -\overline{\beta}^{(1)}_i - K = 0$ for a specific set of pump, signal, and idler modes. By neglecting the coupling between the two waveguides in Eq.~(\ref{NL_sect_EQ}), we implicitly also have neglected the effects of idler loss in Wg-2 on the biphoton evolution in the nonlinear sections.
The solution of Eq.~(\ref{NL_sect_EQ}) for the first nonlinear section NL$_1$ with $0 \leq z \leq  l$ gives the photon pair amplitude originating from the first source. At $z = l$ this is
\begin{equation}
    \boldsymbol{\phi}(l) =
    \begin{bmatrix}
    1 &&
    0
    \end{bmatrix}^\text{T} \phi_0(l)
\end{equation}
with
\begin{equation} \label{SPDCsorce_pair_amplitude}
\phi_0(l) = (A_p(0)  l/2) \cdot e^{-\ii \left(\beta_s+ \beta^{(1)}_i + \frac{\Delta \beta_{NL} }{2} \right)l} \snc\left( \frac{\Delta \beta_{NL} l}{2}\right),
\end{equation}
where $\Delta \beta_{NL} = \beta_p - \beta_s -\beta^{(1)}_i - K$ denotes the phase mismatch for SPDC in NL$_{1}$ and NL$_{2}$.

Within the PT coupler section of length $L$, the evolution of amplitude
$ \boldsymbol{\phi}(z) = \begin{bmatrix}
\phi_{11}(z) & \phi_{12}(z)
\end{bmatrix}^\text{T}$ is governed by
\begin{equation}\label{Coupler_sect_EQ}
\frac{\partial \boldsymbol{\phi}(z)}{\partial z}  = \left(-\ii\beta_s\mathds{1} + \mathbf{M}_i \right)\boldsymbol{\phi}(z),
\end{equation}
where we have used the propagation matrix $ \mathbf{M}_i $ from Eq.~(\ref{PT_evol_matrix}).
We propagate the photon-pair amplitudes generated in NL$_1$ through the coupler section of the interferometer with $l \leq z \leq l+L$ by solving Eq.~(\ref{Coupler_sect_EQ}) making use of the eigenvalues and eigenvectors of matrix $ \left(-\ii\beta_s\mathds{1} + \mathbf{M}_i \right) $. At $z = l+L$, this results in
\begin{equation}
\begin{split}
\boldsymbol{\phi}(l+L) =
 & e^{-\ii \left(\beta_s + \Bar{\beta}_i  \right)L}  e^{- {\gamma_i L}/{2} } \\
    & \times \begin{bmatrix}
    \cosh^2{\Theta}e^{{\xi_i L}/{2}} -  \sinh^2{\Theta}e^{-{\xi_i L}/{2}}\\
    \ii \sinh{\Theta} \cosh{\Theta} ( e^{-{\xi_i L}/{2}} - e^{{\xi_i L}/{2}} )
    \end{bmatrix} \phi_0(l), \label{2Ph_Amp_PTc}
\end{split}
\end{equation}
where $\sinh{\Theta} \equiv (2C_i)/\sqrt{ (\gamma_i + \xi_i - 2\ii \Delta \beta_i)^2 - (2C_i)^2 }$ and $ \cosh^2{\Theta} -  \sinh^2{\Theta} = 1$ (see Appendix~\ref{app:biphoton} for details of derivation). Finally, to obtain the complete biphoton state amplitude at the end of the interferometer, $ z = 2l+L $, we superpose the photon pair amplitudes in Eq.~(\ref{2Ph_Amp_PTc}) with the pair generation amplitude arising from the second source, NL$_2$. This results in the amplitude
\begin{equation}
\begin{split}
\boldsymbol{\phi}(2l+L) = & \begin{bmatrix}
    e^{-\ii \beta^{(1)}_il}&& 0\\
    0 && e^{-\ii \beta^{(2)}_il}
    \end{bmatrix}e^{-\ii \beta_s l} \boldsymbol{\phi}(l+L) \\
    & + \begin{bmatrix}
    1 \\
    0
    \end{bmatrix} \phi_0 e^{-\ii \beta_p (l+L)}, 
\end{split}
\label{eq:amplitudes_final}
\end{equation}
where we note that the second nonlinear source can contribute only to $\phi_{11} $ since both photons are generated in and remain confined to Wg-1 due to this source.

\subsection{Measurable signal photon intensity}
\begin{figure}[tb]
    \centering
    \includegraphics[width = \linewidth]{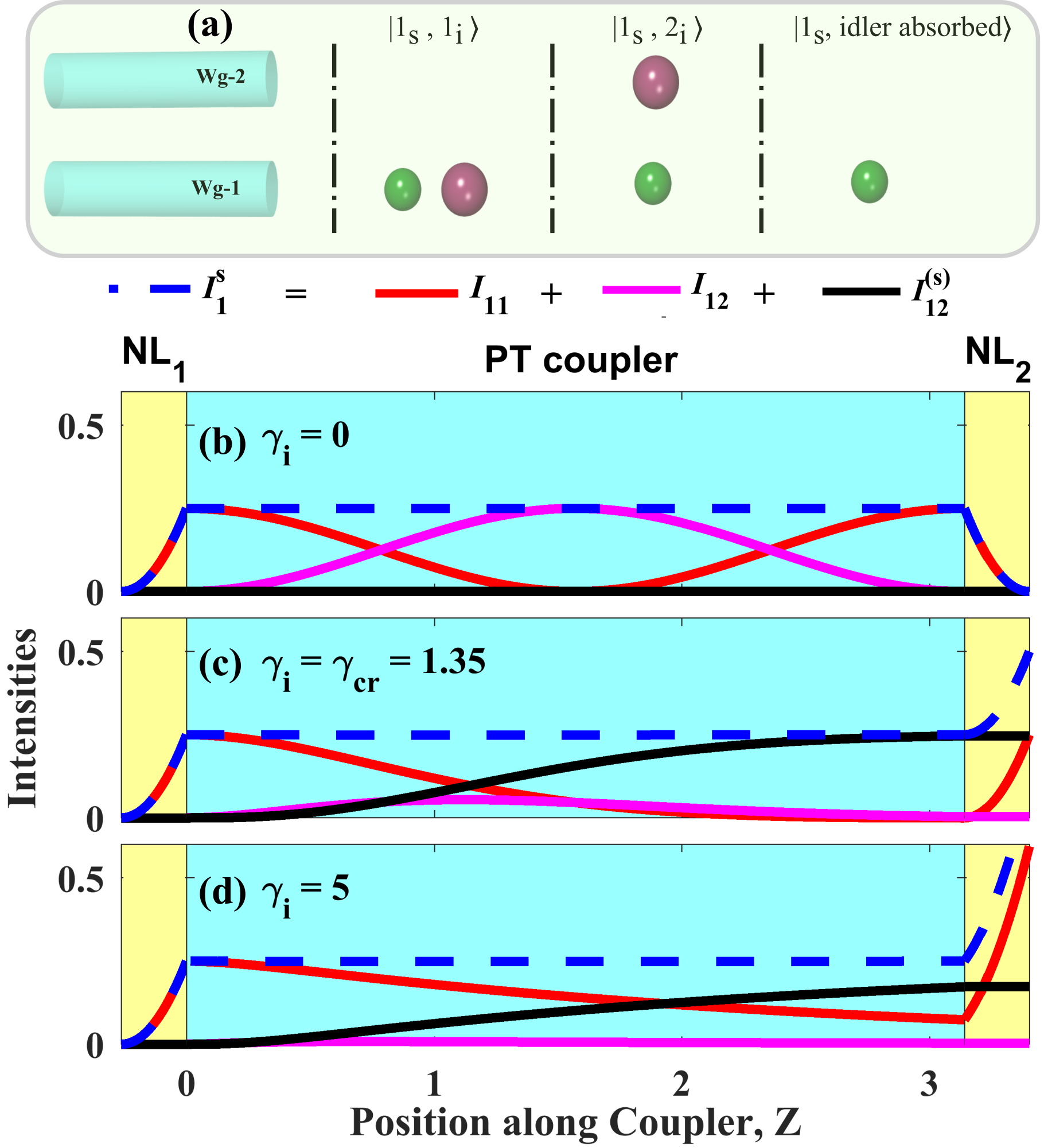}
    \caption{(a) Top panel shows waveguides in which signal and idler photons are present for the biphoton $(I_{11} \& \ I_{12})$ and single photon $I^{(\text{s})}_{12} $ intensity contributions (see text for their description).(b) For lossless PT coupler, biphoton amplitudes from sources NL$_1$ and NL$_2$ interfere destructively. (c) At the critical loss, $\gamma_i = \gamma_{cr}$, pair generation in NL$_1$ and NL$_2$ become totally distinguishable and their amplitudes add up incoherently. (d) In the broken PT regime, nonlinear interferometer exhibits increased indistinguishability and coherence between pair generation amplitudes from the two sources.
    In (b-d), a special coupler design with $C_i L =\pi$ is considered.}
    \label{Int_Evolu}
\end{figure}

Quantum spectroscopy based on nonlinear interference and induced coherence requires only detection of the signal photon intensity. In the implementation discussed here, depicted in Fig.~\ref{schematic}(a), the intensity measurement is performed only on the signal mode in Wg-1, as the signal photon generated in either source will not couple to Wg-2. 
The detection of a signal photon in Wg-1 could result from one of three possibilities for the state of the idler photon shown schematically in Fig.~\ref{Int_Evolu}(a), corresponding to the three probability amplitudes $\phi_{11}$, $\phi_{12}$ and $\phi_{12}^{\text{(s)}}$. These possibilities are distinguishable, and hence the total signal intensity is a sum of the intensities of the three contributions as
\begin{equation}
    I^{\text{s}}_{1} = I_{11} + I_{12} + I^{(\text{s})}_{12}.
\end{equation}
Here, the intensity contributions stemming from the two-photon amplitudes can be directly calculated from Eq.~(\ref{eq:amplitudes_final}) as $I_{11}(z) = |\phi_{11}(z)|^2$, $ I_{12}(z) = |\phi_{12}(z)|^2 $. 

Next, we derive the $ I^{(\text{s})}_{12} $ contribution to the signal intensity.  The probability amplitude for detecting the signal photon at position $z$ with its partner idler photon being absorbed at position $ z' $ is given by $ \phi_{12}^{\text{(s)}}(z, z') $. Following Ref.~\cite{Antonosyan:2014-43845:PRA}, its evolution is given by
\begin{equation} \label{eq:absorbed_amp}
  \frac{\partial \phi^{\text{(s)}}_{12}(z, z')}{\partial z}  =  \begin{cases}
  -\ii\beta_s \ \phi^{\text{(s)}}_{12}(z, z') & \text{for $z' \leq z$} \\
  0 & \text{otherwise}
  \end{cases}
\end{equation}
with the initial condition $ \phi^{\text{(s)}}_{12}(z', z') = -\ii \sqrt{2\gamma_i} \ \phi_{12}(z')$.
The signal intensity $ I^{\text{(s)}}_{12}(z) $ is accumulated over the length of the coupler from the absorbed idler amplitudes $ \phi^{\text{(s)}}_{12}(z, z') $ and is equal to
\begin{align*}
    &I^{\text{(s)}}_{12}(z) = \int^{z}_l |\phi^{\text{(s)}}_{12}(z, z')|^2 dz'= 2\gamma_i\int^{z}_l  |\phi_{12}(z')|^2 dz'; \ \text{for} \ z > l,
\end{align*}
with $ I^{\text{(s)}}_{12}(l) = 0 $. The second nonlinear source does not contribute to this amplitude, since it is assumed to be lossless and not coupled to Wg-2.

Overall, we find that the second nonlinear section can only affect $ I_{11}$.
Only the first source NL$_1$ placed before the coupler can contribute to all three intensities $ I_{11}$, $I_{12}$, and $I^{(\text{s})}_{12}$. Hence, the two sources only interfere, if NL$_1$ contributed to $ I_{11}$ at the end of the structure, since both sources are indistinguishable in this intensity component. This means that induced coherence between the sources can only appear, if there is a non-zero probability for the idler photon from NL$_1$ to be located in Wg-1 in the region NL$_2$. The dynamics of induced coherence is thus dependent on the state of the idler photon generated from NL$_1$, which itself is controlled by the properties of the coupler, i.e its coupling strength, length, and the loss in Wg-2.

\subsection{Constructive and destructive photon interference and its dependence on idler loss} \label{sec:interference}

To fully understand the dependence of biphoton interference 
on the idler loss, we explicitly calculate the intensity stemming from the case where both photons are in Wg-1 using Eq.~(\ref{eq:amplitudes_final}) and  Eq.~(\ref{2Ph_Amp_PTc}), as this is the only contribution to the signal intensity that can show interference. It is
\begin{equation} \label{I_11_final}
  I_{11}(2l+ L)   = \left[ 1+ |V_{\gamma_i}|^2 + 2\RE \left\{ V_{\gamma_i} \times \text{e}^{\ii(\Delta \beta_{NL} + K) (L+l)}  \right\} \right ]|\phi_0|^2 , 
\end{equation}
%

%
%
with
\begin{equation} \label{Eq:V_complex}
    V_{\gamma_i} = e^{\left(\ii\Delta \beta_i L - {\gamma_i L}/{2}\right)}
    \left[\cosh{\frac{\xi_i L}{2}} + \left(1 + 2 \sinh^2{\Theta} \right) \sinh{\frac{\xi_i L}{2}} \right].
\end{equation}

%

In $I_{11}$, biphoton interference is clearly manifested in the last term of Eq.~(\ref{I_11_final}), where the visibility of the resulting interference fringes as a function of $ \Delta \beta_{NL} $ is determined by parameter $V_{\gamma_i}$.
The other contributions to the signal intensity can only result from photon-pair generation in the first nonlinear section NL$_1$ and the explicit expressions for intensities $ I_{12}(2l+L)$ and $ I^{\text{(s)}}_{12}(2l+L) $  do not show any signatures of two photon interference like $ I_{11}(2l+L)$ (see Appendix~\ref{app:signal}). 
As such, the magnitude of $ |V_{\gamma_i}| $ could be regarded as the measure of the indistinguishability of the two photon pair sources in the interferometer enabling biphoton interference. 

We see from Eq.~(\ref{I_11_final}) that whether the two photon interference is constructive or destructive depends on the additional phase acquired by the idler photon as it propagates through the PT coupler. This is encapsulated in $V_{\gamma_i}$, which is defined through the coupler asymmetry, the coupling strength, the length of the coupler, and the idler loss.


For a  symmetric coupler with equal real parts of the idler propagation constants in both waveguides, $\Delta \beta_i = 0$, the expression in  Eq.~(\ref{Eq:V_complex}) 
simplifies to 
\begin{equation} \label{Eq:V_BPT}
     V_{\gamma_i} = e^{-\frac{\gamma_i L}{2}}\left[ \cos{\frac{\sigma_i L}{2}} + \frac{\gamma_i}{\sigma_i} \sin{\frac{\sigma_i L}{2}}\right]  \text{with}\  \sigma_i = \sqrt{(2C_i)^2 - \gamma^2_i}
\end{equation}
below the PT threshold loss, $\gamma_i \leq 2C_i$. Above PT threshold loss $\gamma_i > 2C_i$,  $V_{\gamma_i}$ can be similarly shown to be 
\begin{equation} \label{Eq:V_APT}
    V_{\gamma_i} =  e^{-\frac{\gamma_i L}{2}}\left[ \cosh{\frac{\eta_i L}{2}} + \frac{\gamma_i}{\eta_i} \sinh{\frac{\eta_i L}{2}}\right]  \text{with}\  \eta_i = \sqrt{\gamma^2_i - (2C_i)^2 }.
\end{equation}
We note from Eq.~(\ref{Eq:V_BPT}) and Eq.~(\ref{Eq:V_APT}) that $V_{\gamma_i}$ is a real function and depends on idler loss $\gamma_i$. Further,
only constructive interference is possible in the PT-broken regime, since $V_{\gamma_i} > 0$ according to Eq.~(\ref{Eq:V_APT}). Importantly, the sign of $V_{\gamma_i}$ is unrestricted in the PT-symmetric regime. The transition between constructive and destructive interference can therefore occur at a critical idler loss $\gamma_{cr}$, when
\begin{equation} \label{eq:gamma_cr}
    V_{\gamma_{cr}} = 0 .
\end{equation}
These features can be employed to tailor the interferometer response to loss variations as we discuss below.


\subsection{Evolution of photon intensities}

We show the characteristic evolution of the different intensity contributions along the interferometer in Figs.~\ref{Int_Evolu}(b-d) for three different values of loss in Wg-2. We consider the case of phase matched SPDC with $\Delta\beta_{NL} =0$ and a symmetric PT coupler where $\Delta \beta_i = 0$. The relative phase between pair amplitudes from the two sources stems from the constant phase factor $ K (L+l) $ and the loss-dependent phase of $V_{\gamma_i}$. For simplicity, here we assume that the QPM grating vector is such that $ K (L+l) = 2n\pi \ \text{with} \ n \in \mathbb{N}$, resulting in constructive biphoton interference if no coupling was present between the two Wgs. However, since coupling is present, the phase of $V_{\gamma_i}$ additionally modifies the interference. To maximize the photon-pair interference in the second nonlinear section, we choose the length of the PT coupler equal to twice the coupling length in absence of loss, corresponding to $C_i L =\pi$. The nonlinear section length is taken to be $C_i l = 0.05$, much less than the coupling length. In practice, nonlinear section length $l$ can be always suitably chosen to ensure that $ K (L+l) = 2n\pi$ if the coupler length $ L $ has to be kept constant. In the following numerical simulations, we keep the normalized value of the coupling constant fixed as $C_i = 1$ with no loss of generality. Other parameters such as lengths, propagation constants and losses are assumed to be appropriately scaled by $C_i$ and rendered dimensionless.

We present in Fig.~\ref{Int_Evolu}(b) the Hermitian case in which the PT coupler is lossless with $\gamma_i = 0$. 
In this case, the idler photon of the pair generated in NL$_1$ completely couples to Wg-2 and back such that $|\phi_{11}(l+L)|= |\phi_{11}(l)|$ while $I_{12}(l+L) = 0$. 
The photon-pair amplitude generated in NL$_2$ is coherently added to the amplitude stemming from NL$_1$, which in the considered case leads to complete destructive interference and a signal intensity of zero at the end of the interferometer. This is the manifestation of the fact that in this case the contributions of the two nonlinear sections to the photon-pair amplitude are indistinguishable, enabling interference.
Indeed, for the considered parameters according to Eq.~(\ref{Eq:V_BPT}) the visibility parameter is $V_{\gamma_i} = \cos(C_i L)$, 
leading to $V_{\gamma_i} = -1$ for the chosen coupler length, such that the final biphoton intensity $I_{11}$ becomes zero. Since no loss is present in Wg-2, the signal photon amplitude associated with the absorbed idler photon is also zero in this case (see $ I^{(\text{s})}_{12} $) and thus, the total final signal intensity $I^{\text{s}}_{1}$ vanishes.

Introducing loss in Wg-2 for the idler photon changes the response of the interferometer. We show this in Fig.~\ref{Int_Evolu}(c) for a loss $\gamma_{i}=\gamma_{cr}=1.35$ according to Eq.~(\ref{eq:gamma_cr}), which we refer to as the critical loss. At $\gamma_{cr}$, at the end of the PT coupler, the idler photon of a pair generated in NL$_1$ is either present in Wg-2 or is absorbed in Wg-2 during propagation in the coupler, such that $ I_{11}(l+L) = 0$. Therefore in this case, no interference takes place between the two sources, because they are completely distinguishable and $I_{11}(2l+L)$ only contains contributions from the second source NL$_2$. 
Consequently, the final output signal intensity $ I^s_1(2l+L) $ from the interferometer results from a completely incoherent addition of pair amplitudes originating from the two sources. 

We note that in standard implementations of quantum nonlinear interferometers \cite{Chekhova:2016-104:ADOP}, biphoton amplitudes from individual sources approach complete distinguishability, leading to their fully incoherent addition and the absence of interference, only at infinitely high losses. Due to the interplay of coupling and loss in our structure, complete distinguishability can be reached for the finite loss $\gamma_{cr}$. As we show in the following, the value of $\gamma_{cr}$ can be controlled by appropriately choosing the coupling constant and length of the PT coupler. We also find that even several such critical loss values can exist for a particular structure, where for a fixed coupling strength the coupler length has to be made longer to increase the number of critical points corresponding to complete distinguishability (see Appendix~\ref{app:signal}).

The behaviour of the interferometer is quite different for losses above the PT symmetry breaking threshold $\gamma_{th} = 2C_i$, since the points of critical loss exist only in the PT symmetric regime as discussed in Sec.~\ref{sec:interference}. 
This aspect is reflected in Fig.~\ref{Int_Evolu}(d), where we plot the various intensity contributions at a loss magnitude of $\gamma_i = 5$. Here, the final signal intensity $ I_{1}^s (2l + L) $ as well as $ I_{11}(2l + L) $
are larger compared to Fig.~\ref{Int_Evolu}(c) despite stronger loss. This apparently counter-intuitive result can be explained by considering the propagation dynamics of the idler photon in the PT coupler for different losses. 
The nature of the two supermodes $ \mathbf{V}_\pm $ and their corresponding eigenvalues dictate the idler photon dynamics as expressed by Eq.~(\ref{Formal_sol}). As discussed before, these  eigenmodes of the PT coupler remain symmetric in the two Wgs below $ \gamma_{th}$ and exhibit equal losses as is shown in Fig.~\ref{schematic}(b). 
For losses above $\gamma_{th}$, this symmetry is broken and the two Wgs start to effectively decouple from each other. The $\mathbf{V}_+$ eigenmode, with its corresponding eigenvalue shown by the red solid curve in Figs.~\ref{schematic}(b, c) tends to be localized in the lossless Wg-1, while the other eigenmode localizes in the lossy Wg-2. As a result, the $\mathbf{V}_+$ mode experiences progressively lower effective losses while the  $\mathbf{V}_-$ mode becomes increasing lossy in the broken PT regime. Since the first source NL$_1$ is positioned in Wg-1 in the interferometer, its biphoton amplitude favourably excites the $\mathbf{V}_+$ idler supermode in the broken PT regime and hence, the idler may experiences lower effective losses for $\gamma > \gamma_{th}$. As this supermode features a high probability for the idler photon to be localized in Wg-1, this means that also $ I_{11}(l + L) $ is increased for $\gamma > \gamma_{th}$, which can interfere with the contribution from NL$_2$. 
Hence, the indistinguishability of pair generation between the two sources is improved as the idler loss strength increases in Wg-2 in the broken PT regime.

\subsection{Signal fringes and sharp transition at critical loss}

After discussing the signal intensity evolution within the structure for several specific values of idler loss and a fixed phase mismatch, we now explicitly demonstrate the sharp change in the behaviour of the interferometer around the critical loss magnitude for a varying mismatch $\Delta \beta_{NL}$. In experimental realizations, this mismatch usually corresponds to the wavelength of the detected signal mode for a fixed pump wavelength. 

\begin{figure}
    \centering
    \includegraphics[width = \linewidth]{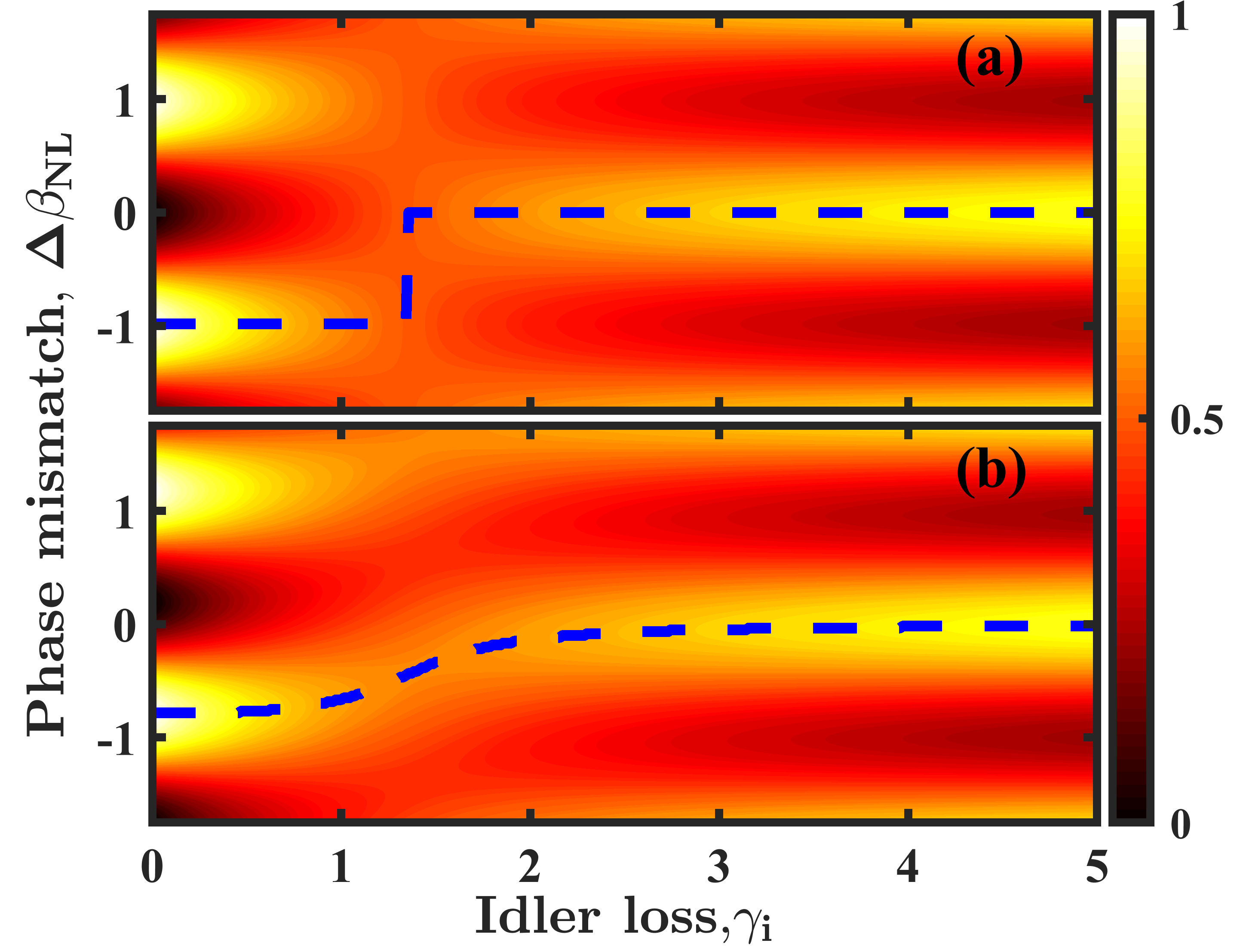}
    \caption{Normalized total signal intensity $I^{\text{s}}_{1} $ vs. the idler absorption strength ($\gamma_i$) and phase mismatch parameter ($\Delta \beta_{NL}$).  (a)~In a symmetric coupler, $\beta^{(2)}_i = \beta^{(1)}_i$, signal intensity fringe position exhibit a sharp shift by half a period at the critical loss $\gamma_i = \gamma_{cr}$. 
    (b)~For the asymmetric coupler with $ \Delta \beta_i = 0.2 $, signal fringe position shifts gradually with the absorption strength. For all the plots, $C_i L = \pi$ and $C_i = 1$.}
    \label{Fringe_Shift}
\end{figure}

We show in Fig.~\ref{Fringe_Shift}(a) the dependence of the total signal intensity $I^{\text{s}}_{1}$ on the idler loss strength $\gamma_i$ and the phase mismatch parameter $\Delta \beta_{NL}$. We clearly observe the fringes as a function of $\Delta \beta_{NL}$. 
For zero loss, the visibility of interference is perfect, being equal to unity, and the interference maxima appear for specific $\Delta \beta_{NL}$. 
As $\gamma_i$ increases, the visibility is reduced and it completely vanishes exactly at the critical loss $\gamma_{cr} = \text{1.35}$. 

Strikingly, the signal intensity fringe exhibits a sharp shift by half a period as the idler loss strength changes across this critical loss magnitude. This is illustrated by the blue dashed line overlaid on the signal intensity contour plot that tracks the position of the central intensity maximum. We also see that such behaviour does not occur with increasing idler losses beyond the PT threshold loss $\gamma_{th} = 2$. In fact, the signal intensity fringe acquires a specific position in $\Delta \beta$ and its visibility improves with increasing idler loss.

We have, thus far, only visualised the case of symmetric coupler where the two Wgs have equal propagation constants for the idler mode, i.e., $\beta^{(2)}_i = \beta^{(1)}_i$ and thus $\Delta \beta_i = 0$. 
We now analyse the effect of breaking this symmetry when $\Delta \beta_i \ne 0$.
The resulting signal intensity behaviour is displayed in Fig.~\ref{Fringe_Shift}(b) for a detuning magnitude of $\Delta \beta_i = 0.2$. Unlike the PT symmetric case, the interferometer employing the asymmetric coupler shows a gradual shift in the position of the signal intensity fringe with a change in the idler loss strength. Also, the visibility of the interference does not reduce to zero for any specific loss value.

\begin{figure}
    \centering
    \includegraphics[width = \linewidth]{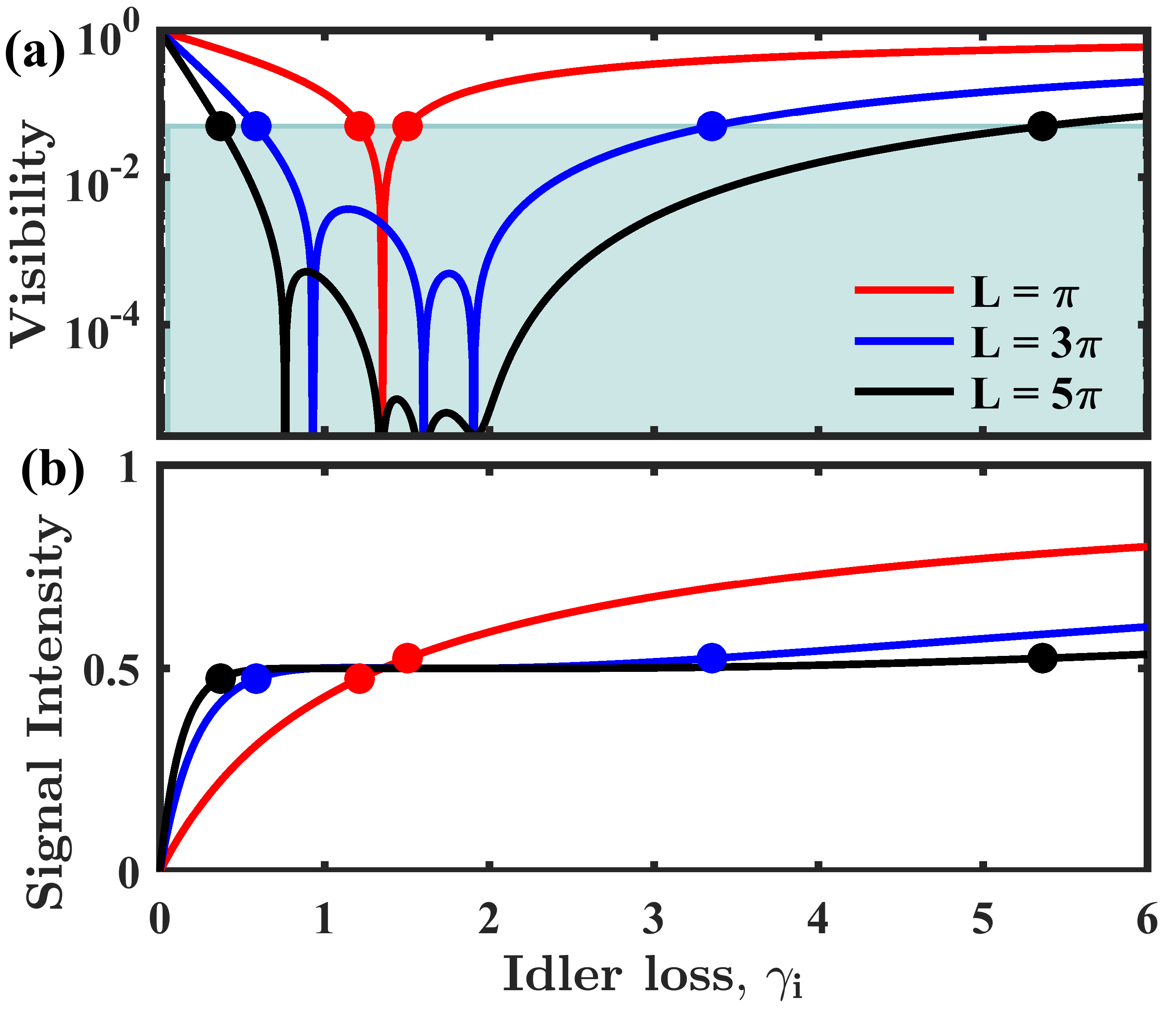}
    \caption{(a) Visibility ${\cal V}$ of the interference in the signal intensity at the end of the nonlinear interferometer containing the PT coupler vs. the idler loss for different coupler lengths. (b)~Corresponding normalized total signal intensity at $\Delta\beta_{NL}=0$. The dots denote the boundaries of regions where the visibility is below 0.05 and the intensity is close to 0.5.}
    \label{Visibility_CoupLen}
\end{figure}

\subsection{Sensing of idler losses}

Based on the nontrivial dependence of the signal intensity on the idler loss, next we suggest a specific sensing application of our scheme. We have shown the dependence of the interference fringes on the loss in Fig.~\ref{Fringe_Shift}(a) and we explicitly show the associated variation of their visibility in Fig.~\ref{Visibility_CoupLen}(a) with the red curve. Here, the visibility $ {\cal V} $ of the signal interference fringe for loss $ \gamma_i $  is calculated as $ {\cal V} =  \{\mathrm{max}_{\Delta \beta_{NL}}(I^{\text{s}}_{1}) - \mathrm{min}_{\Delta \beta_{NL}}(I^{\text{s}}_{1})\}/ \{\mathrm{max}_{\Delta \beta_{NL}}(I^{\text{s}}_{1}) + \mathrm{min}_{\Delta \beta_{NL}}(I^{\text{s}}_{1})\}$, where $ I^{\text{s}}_{1} \left(L+2l\right) $ is the signal intensity as a function of the phase mismatch $\Delta \beta_{NL}$ at constant loss magnitude. We note that ${\cal V}$ is different from the visibility parameter  $ V_{\gamma_i} $ discussed before for the $I_{11}$ intensity due to the contributions of $I_{12}$ and $I^{(\text{s})}_{12}$ to the total signal intensity.

In Fig.~\ref{Visibility_CoupLen}(a), we mark a region in grey where the visibility ${\cal V}$ is below 0.05. This value can serve as a specific threshold magnitude of visibility below which interference fringes can not be detected reliably. We have highlighted the corresponding two points on the red curve between which the visibility is below this threshold value. The critical loss $\gamma_{cr}$, for which the visibility vanishes completely, lies between these two points. In Fig.~\ref{Visibility_CoupLen}(b) we show the corresponding signal intensity $I^{\text{s}}_{1} (\Delta \beta_{NL} = 0)$ for the phase matched case along with the two threshold visibility points. In the intermediate region between these points we have $I^{\text{s}}_{1} \approx 0.5 \times I^{\text{s}}_{1} (\gamma_i \to\infty)$ (see Fig.~\ref{Visibility_CoupLen}(b)), signifying that the two sources of photon pairs are almost distinguishable. Hence, no notable interference pattern in the signal intensity can be observed. 

The range of loss magnitudes where the two photon-pair sources are nearly completely distinguishable and the visibility is very close to zero can be controlled by adapting the design parameters of the nonlinear interferometer. We show this by considering designs with couplers of lengths $ L = 3\pi $ and $5\pi$, shown in Figs.~\ref{Visibility_CoupLen}(a, b) with the blue and black curves, respectively. The visibility plots clearly show that a larger loss region with visibilities below 0.05 can be realized with several critical loss points lying within this region. It is important to realize that all such critical points lie below the PT-symmetry-breaking threshold $\gamma_{th} = 2$. Furthermore, for loss magnitudes within this region, the signal intensity stays almost constant, regardless of the precise value of the loss (see Appendix~\ref{app:signal}). 

The peculiar features discussed above could be applicable to sensing scenarios that require to determine whether the concentration of the analyte responsible for the loss is within a specific target range. Importantly, the boundaries of the range with constant signal intensity can be fixed by appropriately choosing the length and coupling strength of the PT coupler. Furthermore, the suggested device can be preferably used to sense very high absorption. Here, contrary to standard nonlinear interferometers, the visibility of interference is revived and increasing for larger losses, which renders their detection easier.
\subsection{Broadband absorption spectroscopy}

Spectroscopic applications employing nonlinear interferometers typically aim to achieve operation in a broad spectral range such that the frequency dependent absorption profile of an analyte can be measured around the idler photon frequency by performing detection at the corresponding frequencies of the signal photon. Such operation is essentially achieved by employing broadband SPDC in the interferometer where two short nonlinear sources of photon pairs are pumped by a continuous wave pump laser at a fixed frequency.
We now perform a similar frequency dependent loss analysis for the interferometric scheme that we propose in this work. 

To illustrate the spectral response, we start by considering the frequency dependence of the phase mismatch parameter $ \Delta \beta_{NL} $ due to the dispersion of signal and idler modes. Under a first order approximation, the propagation constants of the signal and idler modes can be described as  $ \beta^{(1)}_{s, i}(\overline{\omega}_{s,i} + \Delta \omega_{s,i}) =  \beta^{(1)}_{s,i}(\overline{\omega}_{s,i}) +  \left({1}/{v_g} \right)_{s, i} \times ( \Delta \omega_{s,i})$, where $\overline\omega_{s,i}$ are the respective central frequencies and $ v_g $ denotes the group velocity at these frequencies. Assuming a continuous wave pump at frequency $ \omega_{p} = \overline\omega_{s} + \overline\omega_{i} $, the phase mismatch parameter then becomes
\begin{equation} \label{disper_eq}
    \Delta \beta_{NL} = \Delta \left( \frac{1}{v_g} \right) \times \frac{\Delta \omega_{s}}{\omega_p},
\end{equation}
with $ \Delta \left( \frac{1}{v_g} \right) =   \left[ \left({1}/{v_g} \right)_{s} - \left({1}/{v_g} \right)_{i}\right] \times \omega_p$ being the normalized inverse group velocity difference. Here, we have made use of the SPDC phase matching relation and the energy conservation condition $\Delta \omega_{s} + \Delta \omega_{i} = 0 $. 

\begin{figure}
    \centering
    \includegraphics[width = \linewidth]{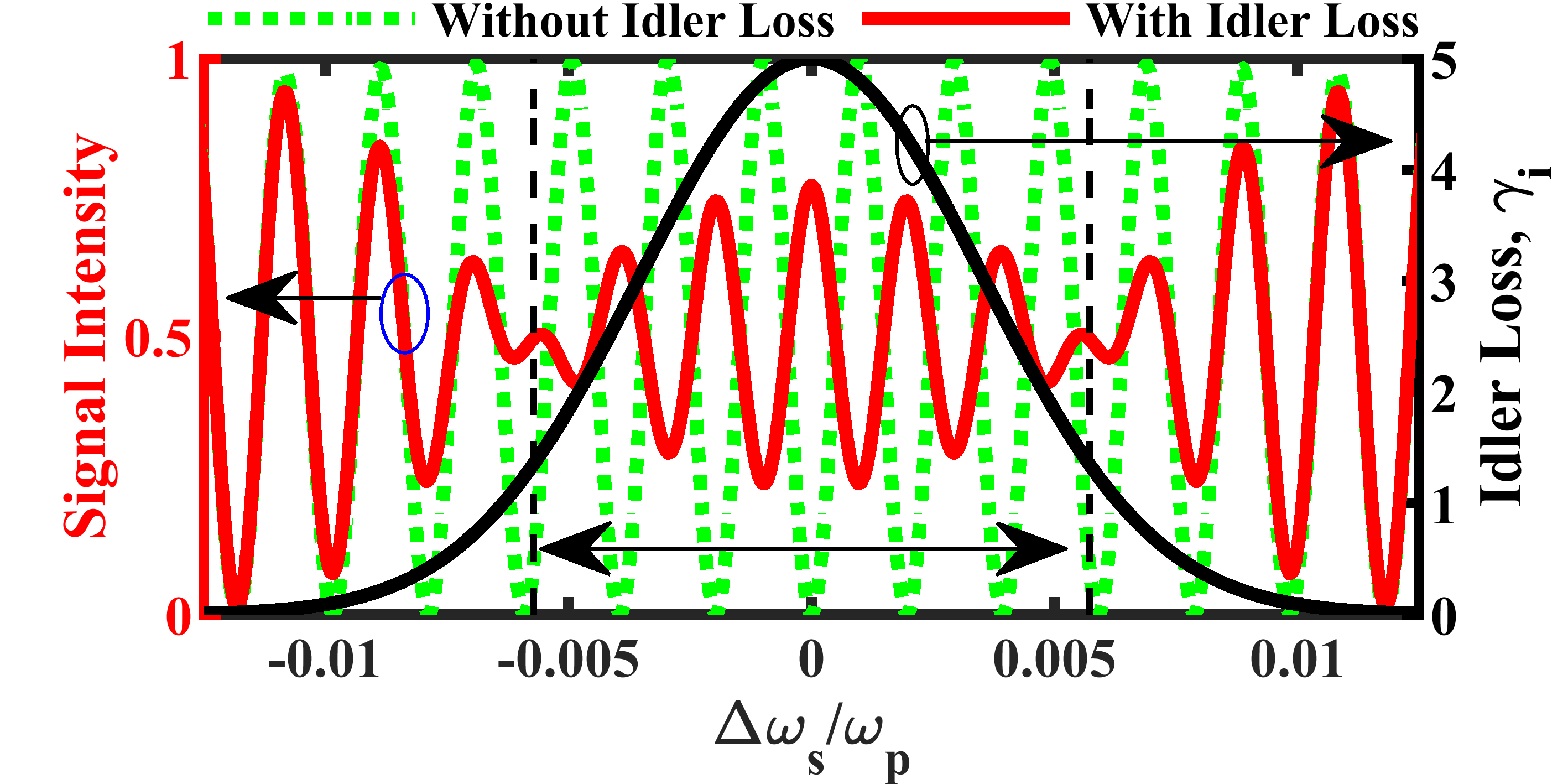}
    \caption{
    Signal spectral intensity (red solid curve) from nonlinear interferometer in presence of a spectrally localized idler absorption profile (black solid curve) with the corresponding frequency given by $ \Delta \omega_i  = - \Delta \omega_s $. Comparison with the reference case of lossless idler, shown by green dotted curve, reveals that nature of interference, constructive or destructive, switches at the critical idler loss $\gamma_i = \gamma_{cr}$ represented by the two vertical dashed lines.}
    \label{Spectral_Loss}
\end{figure}

Bandwidth of the interferometer is limited by the bandwidth of the biphoton amplitude $\phi_0$ from SPDC sources and is given by $\Delta \beta_{NL} = {4\pi}/{l}$ according to Eq.~(\ref{I_11_final}) and Eq.~(\ref{SPDCsorce_pair_amplitude}). This determines the width of the envelope containing the signal intensity fringes.  On the other hand the periodicity of these fringes is determined by the length of the central linear section according to Eq.~(\ref{I_11_final}) and is given by  $\Delta \beta_{NL} = {2\pi}/{(L+l)}$. To detect change in the spectral position and nature of fringes in response to idler loss, densely packed spectral fringes are advantageous, and this regime is realized for a long coupler length $L$ and larger group velocity mismatch $ \Delta\left( \frac{1}{v_g} \right) $ according to Eq.~(\ref{disper_eq}). At the same time though, a large group velocity mismatch also reduces the SPDC bandwidth and therefore, judiciously chosen small length of the SPDC sources ensure that the overall fringe envelope is still broad enough to not alter the relative peak intensities of the signal fringes significantly near the center of the envelope.                        

We demonstrate a typical frequency response of the proposed interferometer by considering a spectrally localized loss profile for the idler mode centered around $\overline\omega_{i}$, as shown by the black solid curve in Fig.~\ref{Spectral_Loss}. Due to the energy conservation, $ \Delta \omega_i  = - \Delta \omega_s$ in the plot. The signal intensity fringes in presence of this loss are, calculated by assuming inverse group velocity difference of $ \Delta\left( \frac{1}{v_g} \right) = 10^3 $, are shown by the red solid curve. We also display the signal intensity for the lossless idler case by the green dotted curve, which serves as a reference. 

The broadband nature of SPDC in the interferometer ensures that the envelope of the signal spectral fringes is much wider than the spectral width of the absorption profile under investigation. Thus, the peaks in the normalized signal intensity shown by the green dotted curve in Fig.~\ref{Spectral_Loss} are close to unity over the width of the black solid curve representing the idler loss profile. Meanwhile, the period of the signal fringes is much smaller than the absorption width such that several fringes encompass the spectral range of loss. 

We see from the red curve in Fig.~\ref{Spectral_Loss} that the nature of interference, either constructive or destructive, and accordingly the final signal intensity, depends on the strength of the idler loss at the corresponding idler frequency. To demonstrate this, we mark the central region of the idler loss profile by two vertical black dashed lines, where the magnitude of loss is greater than the critical loss $\gamma_{cr}$. In this frequency range, maxima (and minima) in signal intensity get reversed as compared to the reference signal intensity due to reversal of the nature of biphoton interference. On the other hand, for frequency ranges in which the idler loss is less than the critical loss, the spectral positions of the signal intensity maxima (and minima) follow that of the case with no loss. In both these cases, the magnitude of fringe contrast which determines the prominence of the peaks in signal intensity depends on the magnitude of the idler loss. In both the central spectral region of higher losses as well as the low-loss region outside it, fringe contrast is significant as long as the loss magnitude is appreciably different from the  critical loss $\gamma_{cr}$.  

\section{Conclusions}

In our work, we investigated sensing with an integrated nonlinear quantum interferometer consisting of a PT coupler between two identical sources of signal and idler photon pairs. We established that a peculiar structure of the eigenmodes in the PT coupler 
enables the tailoring of the quantum interference effects to facilitate the sensing of losses in the second waveguide exposed to an analyte under investigation. In particular, the first-order interference between the signal photons generated in the two nonlinear sources strongly depends on the idler loss in the coupled waveguide, which the signal photons never enter. 

We identify a new effect of a sharp shift of the signal interference fringes in the vicinity of critical loss magnitudes, 
a phenomenon that is strikingly different from the previously studied nonlinear interferometers or PT-symmetric structures.
%
%
Such a peculiar dependence of the 
signal intensity on the idler loss can benefit the sensing applications. 
For instance, if the loss 
depends on the analyte concentration, the interferometer can be engineered such that the signal intensity is constant for a particular allowed range of concentrations and only changes its value when the concentration leaves this target corridor. Furthermore, broadband spectroscopic information about the analyte absorption can be obtained from the dependencies of the signal interference fringes on the wavelengths of the signal and idler photons, where a sudden half-period shift of fringes can directly indicate the change of absorption beyond a critical magnitude. 

We anticipate that these results will stimulate advances in the fundamentals and applications of quantum enhanced sensing with nonlinear interferometers.

\appendix
\newcounter{MainEquations}
\setcounter{MainEquations}{\value{equation}}
\renewcommand{\theequation}{A$\the\numexpr\value{equation}-\the\numexpr\value{MainEquations}$}
\renewcommand{\thesection}{\Alph{section}}
\titleformat{\section}[runin]
  {\large\sffamily\bfseries}
  {APPENDIX \thesection: }
  {0.0em}
  {\MakeUppercase{#1}}
  [\\]
\section{Evolution of biphoton amplitudes in the interferometer} \label{app:biphoton}

Here we provide calculations for the biphoton state amplitudes. The nonlinear interferometer is composed of two waveguides (Wg-1 and Wg-2) that are selectively coupled at the longer wavelength idler radiation through coupling constant $C_i$ as shown in Fig.~\ref{schematic}(a) of the main text. Throughout these calculations we assume that the classical pump radiation and signal photon remain confined to Wg-1 due to their shorter wavelength. Two short sections each of length $l$ at either ends of Wg-1 act as nonlinear elements in the interferometer where pair generation could take place through SPDC of the pump photon. The nonlinearity of these sections could be thought of as being "turned on" by appropriate choice of quasi-phase matching (QPM) poling period. The central section of length $L$ between these two nonlinear sections in  Wg-1 is assumed to behave as a purely linear element. Further, Wg-2 also acts as a linear element.

We describe evolution of biphoton amplitudes in the interferometer in stages. These are as follows: 
\begin{enumerate}
\item 
Firstly, we calculate photon pair amplitudes from the first source NL$_1$  in region $0 \leq z \leq l$ assuming a QPM assisted SPDC process. We simplify this calculation by ignoring the coupling between Wgs. This is justified as long as the nonlinear length $l$ is small enough compared to coupling length of the coupler. This also ensures that we can neglect the effect of loss $\gamma_i$ in Wg-2 on pair generation form NL$_1$. 
\item 
We then propagate the biphoton amplitudes through the passive PT coupler in region $ l \leq z \leq l+L $. In this calculation we include the coupling between Wgs and the loss in Wg-2. 
\item 
Finally, in the region $ l+L \leq z \leq 2l+L $ amplitude for pair generation in NL$_2$ is superposed with the pair amplitude originating in NL$_1$.     
\end{enumerate}

The quantum state in the interferometer is defined by the probability amplitudes $\phi_{11}(z)$, $ \phi_{12}(z) $ and $ \phi^{(s)}_{12}(z, z') $ as described in Sec.~\ref{sec:methods}.

\subsection*{Photon pair amplitude from the first source}
In the nonlinear source section NL$_1$ $(0 \leq z \leq l)$, $\phi_{11}(z)$ evolves as 
\begin{equation}\label{EQA:1}
  \frac{\partial \phi_{11}(z)}{\partial z}  = -\ii \left( \beta_s + \beta^{(1)}_i \right)\phi_{11}(z) + A_p(z), \ \text{with} \ \phi_{11}(0) = 0.
\end{equation}
$ A_{p}(z)$ denotes the classical pump field under the undepleted pump approximation and is given by 
\begin{equation}
\begin{split}
	A_{p}(z) &=  A_p(0)\times\cos{Kz}\times e^{-\ii \beta_p z } \\
	& = \frac{A_p(0)}{2}\left\{e^{\ii K z} + e^{-\ii K z} \right\} e^{-\ii \beta_p z } \\
	& = \frac{A_p(0)}{2}\left\{e^{-\ii(\beta_p - K )z} + e^{-\ii(\beta_p + K )z} \right\}.
\end{split}
\end{equation}
Here $ A_p(0) $ subsumes the strength of quadratic nonlinearity while its effective spatial profile is captured by the factor  $\cos{Kz} $. The grating vector $K = 2\pi/\Lambda$ reflects the QPM that facilitates SPDC in NL$_1$.

By substituting $ \phi_{11}(z) = \tilde{\phi}_{11}(z)e^{-\ii \left(\beta_s+ \beta^{(1)}_i \right)z}$ in Eq.~(\ref{EQA:1}), we get 

\begin{align*} \label{EQA:1.5}
    \frac{\partial \tilde{\phi}_{1,1}(z)}{\partial z}  & = \frac{A_p(0)}{2}\left\{e^{-\ii(\beta_p - K )z} + e^{-\ii(\beta_p + K )z} \right\}e^{\ii \left(\beta_s+ \beta^{(1)}_i \right)z} \\
    & = \frac{A_p(0)}{2}\left\{  e^{-\ii \left( \beta_p - \beta_s -\beta^{(1)}_i - K \right)z} + e^{-\ii \left(\beta_p  - \beta_s -\beta^{(1)}_i + K \right)z} \right\}. 
\end{align*}
The two terms in the above expression denote SPDC processes which are mediated by the $+K$ and $-K$ Fourier components of the nonlinearity. Usually only the term corresponding to $\Delta \beta_{NL} = \beta_p - \beta_s -\beta^{(1)}_i - K$  can be phase-matched such that $  \Delta \beta_{NL} = 0 $ for a specific set of pump, signal and idler modes. Therefore, if we neglect the non phase matched term corresponding to $ e^{-\ii \left(\beta_p  - \beta_s -\beta^{(1)}_i + K \right)z} $, 
\begin{equation}\label{EQA:2}
    \frac{\partial \tilde{\phi}_{1,1}(z)}{\partial z} = \frac{A_p(0)}{2}e^{-\ii\Delta \beta_{NL} z} .
\end{equation}
Solution of Eq.~(\ref{EQA:2})  is 
\begin{equation}
\begin{split}
    \tilde{\phi}_{1,1}(z) - \tilde{\phi}_{1,1}(0) 
    & = \frac{A_p(0)}{2}\frac{\left\{ e^{-\ii\Delta \beta_{NL} z} - 1  \right\}}{-\ii\Delta \beta_{NL}} \\
    & = \frac{A_p(0)\times z }{2} e^{\frac{-\ii\Delta \beta_{NL} z}{2}} \snc \left\{ \frac{\Delta \beta_{NL} z}{2} \right\}, 
\end{split}
\end{equation}
and using $ \tilde{\phi}_{1,1}(0) = 0$,
\begin{equation} \label{EQA:3}
    \phi_{1,1}(z) = \frac{A_p(0)\times z }{2} e^{-\ii \left( \beta_s+ \beta^{(1)}_i + \frac{\Delta \beta_{NL} }{2} \right)z} \snc \left\{ \frac{\Delta \beta_{NL} z}{2} \right\}. 
\end{equation}
Since waveguides are treated as effectively uncoupled in nonlinear section, $0 \leq z \leq l$, and the pump field is present only in Wg-1,  $\phi_{12}(z) $ and $ \phi^{(s)}_{12}(z, z') $ amplitudes are zero in this section of the interferometer. At $z = l$, biphoton amplitudes $ \phi_{11} $ and $ \phi_{12} $ can be denoted as a column vector  
\begin{equation}\label{EQA:3.5}
\boldsymbol{\phi}(l) = 
\begin{bmatrix}
\phi_{11}(l) \\ \phi_{12}(l)
\end{bmatrix} = 
\begin{bmatrix}
1 \\ 0
\end{bmatrix}\phi_0(l),
\end{equation}
with $\phi_0(l)$ being the biphoton amplitude given by  Eq.~(\ref{SPDCsorce_pair_amplitude}) in the main text.
%
%

\subsection*{Photon pair amplitude within the PT coupler}
In the coupler section, $l \leq z \leq l+L$, the evolution of $ \boldsymbol{\phi}(z) $ is given by
\begin{equation} \label{EQA:5} 
\frac{\partial \boldsymbol{\phi}(z)}{\partial z}  = \left(-\ii\beta_s\mathds{1} + \mathbf{M}_i \right)\boldsymbol{\phi}(z),
\end{equation}
with
\begin{equation*}
-\ii\beta_s\mathds{1} + \mathbf{M}_i  = 
\begin{bmatrix}
-\ii \beta_s - \ii(\Bar{\beta}_i + \Delta \beta_i) & -\ii C_i \\
-\ii C_i &  -\ii \beta_s -\ii(\Bar{\beta}_i - \Delta \beta_i) - \gamma_i  
\end{bmatrix}.
\end{equation*}

The eigenvalues $ \overline{\lambda}$ and eigenvectors $\mathbf{V}$ of the propagation matrix $ (-\ii\beta_s\mathds{1} + \mathbf{M}_i)  $ are obtained by first solving the characteristic polynomial $\mathrm{det}(-\ii\beta_s\mathds{1} + \mathbf{M}_i - \overline{\lambda}\mathds{1}) = 0$ to get the two eigenvalues 
$\overline{\lambda}_\pm$ and then solving the corresponding eigenvector equations $ (-\ii\beta_s\mathds{1} + \mathbf{M}_i - \overline{\lambda}_\pm\mathds{1})\mathbf{V}_\pm = 0 $. Explicitly,    
\begin{equation*}
\begin{split}
    \overline{\lambda}_\pm & = -\ii( \beta_s + \Bar{\beta}_i)  + \Tilde{\lambda}_\pm  
\end{split}
\end{equation*}
with $ \Tilde{\lambda}_\pm $ and $ \mathbf{V}_\pm $ given for the PT coupler in Eq.~(\ref{eigen_val_vect}).
Evolution of a general state $ \boldsymbol{\phi}(z) $ in the coupler section can be expressed using  $\overline{\lambda}_\pm$ and $ \mathbf{V}_\pm  $ as 
\begin{equation} \label{EQA:6}
    \boldsymbol{\phi}(z) = \alpha_+e^{\overline{\lambda}_+(z-l)}\mathbf{V_+}  +    \alpha_-e^{\overline{\lambda}_-(z-l)}\mathbf{V_-}.
\end{equation} 
The coefficients $ \alpha_\pm $  are calculated by considering the initial state $ \boldsymbol{\phi}(l) $ at the input of the coupler given in Eq.~(\ref{EQA:3.5}). These are the solution of
\begin{center}
    $ \alpha_+\mathbf{V_+}  +    \alpha_-\mathbf{V_-} =  \begin{bmatrix}
    1 &&
    0
    \end{bmatrix}^\text{T} \phi_0(l)$.  
\end{center}
A straightforward calculation gives 
\begin{equation}
    \alpha_+ = \frac{ \left( \gamma_i + \xi_i - 2\ii \Delta \beta_i \right )  \phi_0(l)  }{\left\{  \left( \gamma_i + \xi_i - 2\ii \Delta \beta_i \right)^2 + \left( 2\ii C_i \right)^2 \right\}},
\end{equation}
and
\begin{equation}
    \alpha_- = \frac{ \left( 2\ii C_i \right )  \phi_0(l)  }{\left\{  \left( \gamma_i + \xi_i - 2\ii \Delta \beta_i \right)^2 + \left( 2\ii C_i \right)^2 \right\}}. 
\end{equation}
Putting these back into Eq.~(\ref{EQA:6}) and after some simplification we get 
\begin{equation} \label{EQA:7}
\begin{split}
 \boldsymbol{\phi}(z) & = e^{-\ii \left(\beta_s + \Bar{\beta}_i  \right)(z-l)}  e^{- {\gamma_i (z-l)}/{2} } \\
   & \times 
    \begin{bmatrix}
    \cosh^2{\Theta}e^{{\xi_i (z-l)}/{2}} -  \sinh^2{\Theta}e^{-{\xi_i (z-l)}/{2}}\\
    \ii \sinh{\Theta} \cosh{\Theta} \left( e^{-{\xi_i (z-l)}/{2}} - e^{{\xi_i (z-l)}/{2}} \right)
    \end{bmatrix} \phi_0(l),
\end{split}
\end{equation} 
where $\sinh{\Theta} \equiv (2C_i)/\sqrt{ (\gamma_i + \xi_i - 2\ii \Delta \beta_i)^2 - (2C_i)^2 }$ and $ \cosh^2{\Theta} - \sinh^2{\Theta} = 1$.

Due to presence of loss in the coupler section, the amplitude $\phi^{(s)}_{12}(z, z')$, which describes the contribution from pairs with lost idler photon, becomes non-zero. Its evolution is given in Eq.~(\ref{eq:absorbed_amp}) with the solution, 
%
%
%
%
%
\begin{equation} 
\phi^{\text{(s)}}_{12}(z, z') = \left\{ -\ii \sqrt{2\gamma_i} \times \phi_{12}(z')\right\} e^{-\ii \beta_s  (z-l)}; \ \  l \leq z' \leq l + L \end{equation}
where $ \phi_{12}(z') $ is given by Eq.~(\ref{EQA:7}) in the coupler.

\subsection*{Photon pair amplitude in the second nonlinear section} 
The amplitudes for photon pair generation in NL$_2$ is easily obtained by using Eq.~(\ref{EQA:3.5}). It is given by  $  \boldsymbol{\phi}(l) \times e^{-\ii \beta_p (l+L)}$. Here, the phase factor $ e^{-\ii \beta_p (l+L)} $ accounts for the evolution of the pump amplitude in the first nonlinear source and the coupler section of Wg-1.  Now, to obtain the final photon pair amplitudes from the interferometer as a whole, we superpose this biphoton amplitude from NL$_2$ with that stemming from the NL$_1$, $  \boldsymbol{\phi}(l+L)$, which is given by Eq.~(\ref{EQA:7}). The resulting expressions for photon pair amplitudes are given by Eq.~(\ref{eq:amplitudes_final}) in the main text.      
%

\section{Signal Intensity from the Interferometer} 
\label{app:signal}

At the output of the interferometer, total signal intensity $I^s_1(2l + L)$ can be calculated by summing up the three different contributions $I_{11}(2l + L) = |\phi_{11}(2l + L)|^2$, $ I_{12}(2l + L) =  |\phi_{12}(2l + L)|^2 $ and $ I^{\text{(s)}}_{12}(2l + L) = \int^{l+L}_l |\phi^{\text{(s)}}_{12}(z, z')|^2 dz'$ $ = 2\gamma_i\int^{l+L}_l  |\phi_{12}(z')|^2 dz'  = 2\gamma_i\int^{l+L}_l I_{12}(z') dz'$. Using Eq.~(\ref{eq:amplitudes_final}),  we obtain
\begin{equation} \label{EQA:I_11_final}
  I_{11}(2l+ L)   = \left[ 1+ |V_{\gamma_i}|^2 + 2\RE \left\{ V_{\gamma_i} \times \text{e}^{\ii(\Delta \beta_{NL} + K) (L+l)}  \right\} \right ]|\phi_0|^2 , 
\end{equation}
with
\begin{equation} \label{EqA:v_vomplex}
    V_{\gamma_i} = e^{\left(\ii\Delta \beta_i L - {\gamma_i L}/{2}\right)}
    \left[\cosh{\frac{\xi_i L}{2}} + \left(1 + 2 \sinh^2{\Theta} \right) \sinh{\frac{\xi_i L}{2}} \right],
\end{equation}
%
%
%
and
\begin{equation} \label{EQA:I_12_final}
    I_{12}(2l+ L)   
    = e^{-\gamma_i L } \left| \sinh{2 \Theta} \times
     \sinh{\frac{\xi_i L}{2}} \right|^2 |\phi_0(l)|^2.
\end{equation}
%
%
%
When the coupler in the interferometer is symmetric ($\Delta \beta_i = 0$),
$\xi_i  = \sqrt{\gamma_i^2 - (2C_i)^2}$. We see that $\xi_i$ is a purely imaginary quantity below the PT threshold and a positive real quantity above the PT threshold. This in turn affects the nature of $V_{\gamma_i}$.      
\subsection*{Below PT threshold} 
Since $\gamma_i < 2C_i$ in PT symmetric regime, $\xi_i$ can be written as $ \xi_i = \ii \sigma_i$, where  $\sigma_i = \sqrt{(2C_i)^2 - \gamma^2_i} > 0$. For substituting $\xi_i$ by $ \ii \sigma_i $ in Eq.~(\ref{EqA:v_vomplex}), we note that $ \cosh{ \frac{\ii \sigma_i L}2} = \cos{\frac{\sigma_i L}{2}}$  and $\sinh{ \frac{\ii \sigma_i L}2} = \ii \sin{\frac{\sigma_i L}{2}}$. Further, $ \left(1 + 2 \sinh^2{\Theta} \right) =  \frac{\gamma_i}{\ii \sigma_i}$. Putting these into Eq.~(\ref{EqA:v_vomplex}) leads to  the Eq.~(\ref{Eq:V_BPT}) for $V_{\gamma_i}$ in the PT symmetric regime.     

\subsection*{Above PT threshold} 
For broken PT regime, we denote $ \eta_i  = \sqrt{\gamma_i^2 - (2C_i)^2} > 0 $. It can be shown that $ \left(1 + 2 \sinh^2{\Theta} \right) = {\gamma_i}/{\eta_i}$ above the PT threshold. Thus, using these definitions in  Eq.~(\ref{EqA:v_vomplex}) readily leads to  Eq.~(\ref{Eq:V_APT}) for $V_{\gamma_i}$.      

\begin{figure} 
    \centering
    \includegraphics[width = 0.85 \linewidth]{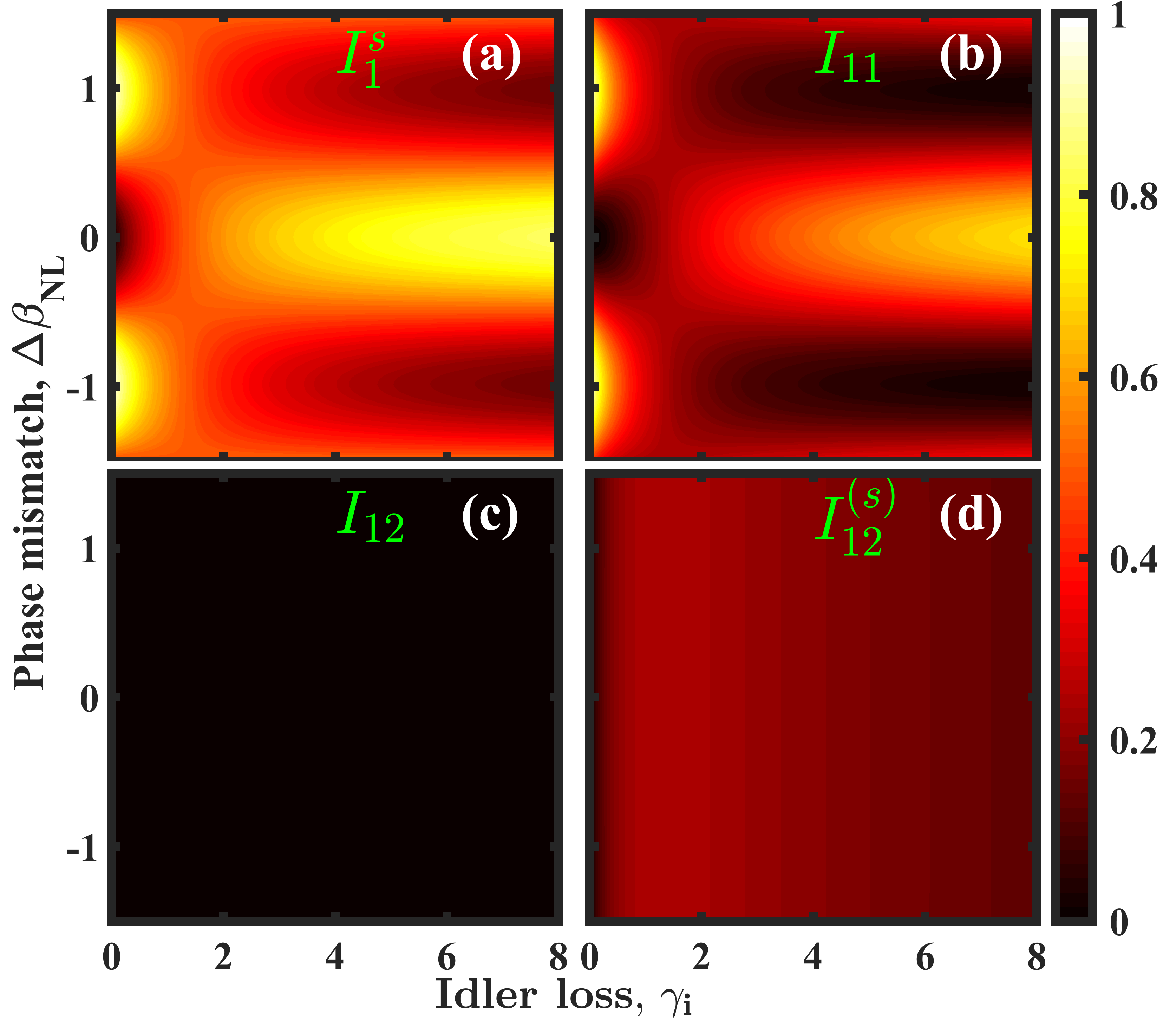}
    \caption{(a) Signal intensity $ I^s_1 $ as a function of phase mismatch $ \Delta \beta_{NL} $ and idler loss $\gamma_i$. (b),(c) and (d) show the three individual contribution to the total signal intensity. $ I_{12} $ and $ I^{\text{(s)}}_{12} $ are devoid of any interference fringes unlike $ I_{11} $. }
    \label{FigA1}
\end{figure}

We show the signal intensity $ I^s_1 $ for the interferometer design presented in the main text ($l = 0.05$, $L = \pi$ and $C_i = 1$) as a function of phase mismatch $ \Delta \beta_{NL} $ and idler loss $\gamma_i$ in Fig.~\ref{FigA1}(a). The three contribution $ I_{11} $, $ I_{12} $ and $ I^{\text{(s)}}_{12} $ are presented in Figs.~\ref{FigA1}(b), (c) and (d) respectively. We see that the $I_{11} $ signal intensity component shows the interference fringes whereas $ I_{12} $ and $ I^{\text{(s)}}_{12} $ add a background which is almost independent of $ \Delta \beta_{NL} $ for any given loss magnitude $\gamma_i$. This means that these two components reduce the overall visibility of fringes in $ I^s_1 $. Similar to the case considered in the main text, we assume $K (L+l) = 2n\pi$ for the relative phase between biphoton amplitudes from the two sources.

\begin{figure} 
    \centering
    \includegraphics[width = \linewidth]{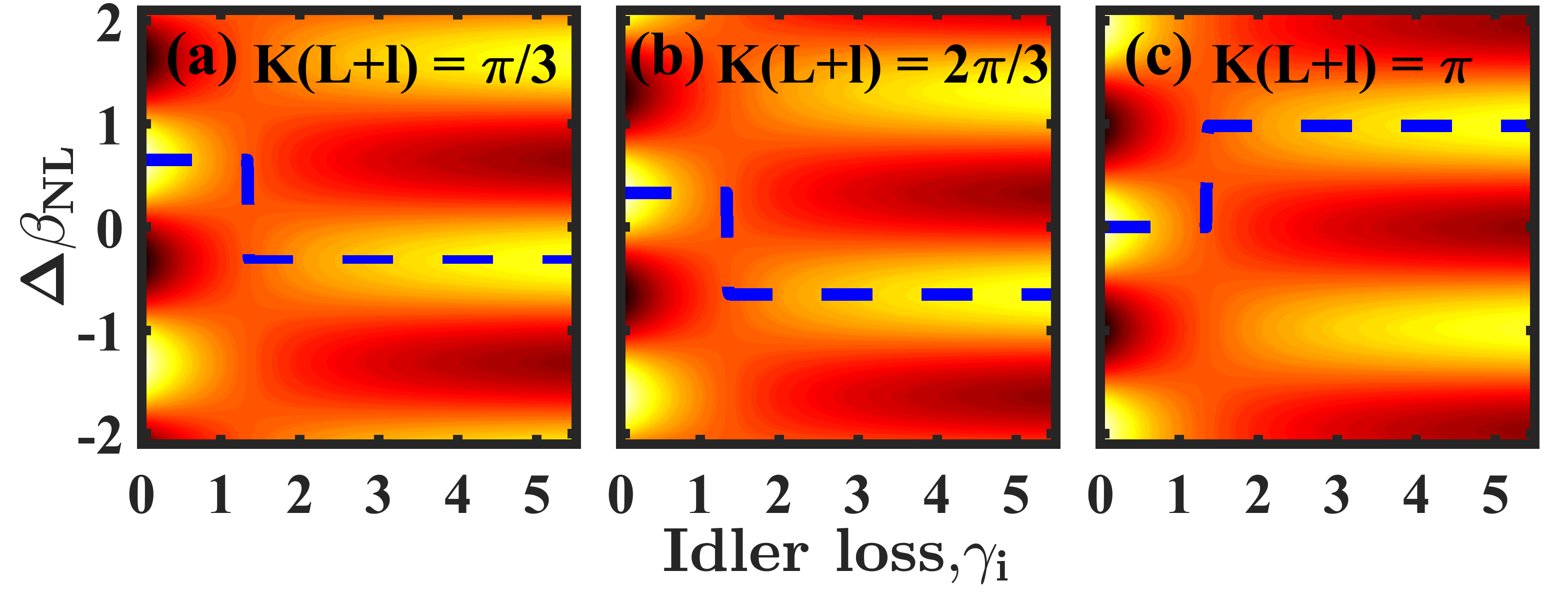}
    \caption{Signal intensity fringes for different relative phases $K(L+l)$ in the interferometer. The feature of fringe shift occurs in each case at the same idler loss value irrespective of $K(L+l)$ phases.}
    \label{FigApped_relPhase}
\end{figure}

The feature of sharp shift in signal intensity fringe at critical idler loss is independent of the exact value of $K (L+l)$. This is relevant for practical interferometer designs where it may not be straightforward to exactly control this relative phase. We show through  Fig.~\ref{FigApped_relPhase}(a)-(c) that the role of  $K (L+l)$ phase is to determine the absolute position of the signal fringes with respect to $ \Delta \beta_{NL}$. In each case, the shift of signal fringes by half a period persists at the same $\gamma_{cr}$ irrespective of the value of $K (L+l)$ phase.    

\begin{figure} 
    \centering
    \includegraphics[width = 0.85\linewidth]{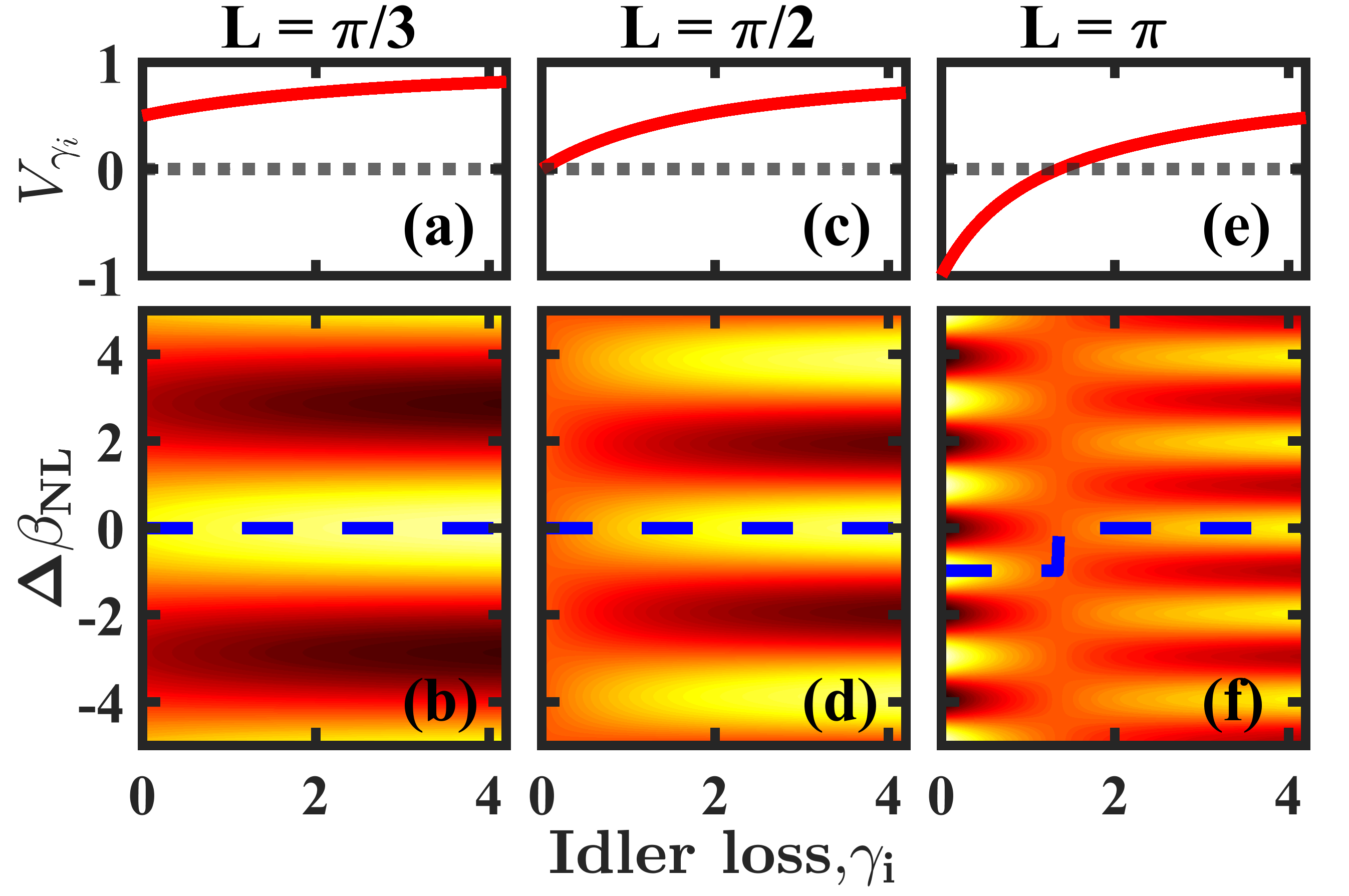}
    \caption{Illustration of the condition for existence of critical idler loss, $\gamma_{cr}$ is $C_i L > \pi /2$, for different $L$ values as indicated by labels. (a,d) When $C_i L \leq \pi /2$, no critical loss points are observed and hence no shift in the fringes. (e,f)~For $C_i L = \pi$, a fringe shift is observed. (a,c,e)~Fringe visibility, (b,d,f)~signal intensity fringes. For all the plots $C_i = 1$. }
    \label{FigApped_ShiftCond}
\end{figure}

We now discuss the condition for existence of critical idler loss points given by Eq.~(\ref{eq:gamma_cr}). Since such points can only exist below the PT threshold, according to Eq.~(\ref{Eq:V_BPT}) $V_{\gamma_i} = 0$ when 
\begin{equation} \label{gamma_critik_eqn}
    \tan{\frac{\sigma_i L}{2}} = - \frac{\sigma_i}{\gamma_i}. 
\end{equation}
There can exist several solutions of Eq.~(\ref{gamma_critik_eqn}), depending on the value of $C_iL$. However, in order to have at least one critical idler loss point, the coupling length at zero loss must be such that $C_i L > \pi/2$. To illustrate this, we plot in Fig.~\ref{FigApped_ShiftCond}, $V_{\gamma_i} $ for three different lengths $L$ keeping $C_i = 1$ fixed. The corresponding signal intensity fringes of the interferometer are also shown in the figure. Figs.~\ref{FigApped_ShiftCond}(a) and~(b) show that when $C_iL = \pi/3$,  $V_{\gamma_i} > 0$ for all idler losses and signal intensity fringes do not show any shift. For $C_iL = \pi/2$, $V_{\gamma_i} = 0$ at zero idler loss (see Fig.~\ref{FigApped_ShiftCond}(c)) and the visibility of signal fringes is also zero at zero loss as shown by Fig.~\ref{FigApped_ShiftCond}(d). When $C_iL > \pi/2$, as considered in Figs.~\ref{FigApped_ShiftCond}(e) and~(f), there exists a critical idler loss point and the associated shift in signal intensity fringe.                        

\begin{figure}[th]
    \centering
    \includegraphics[width = \linewidth]{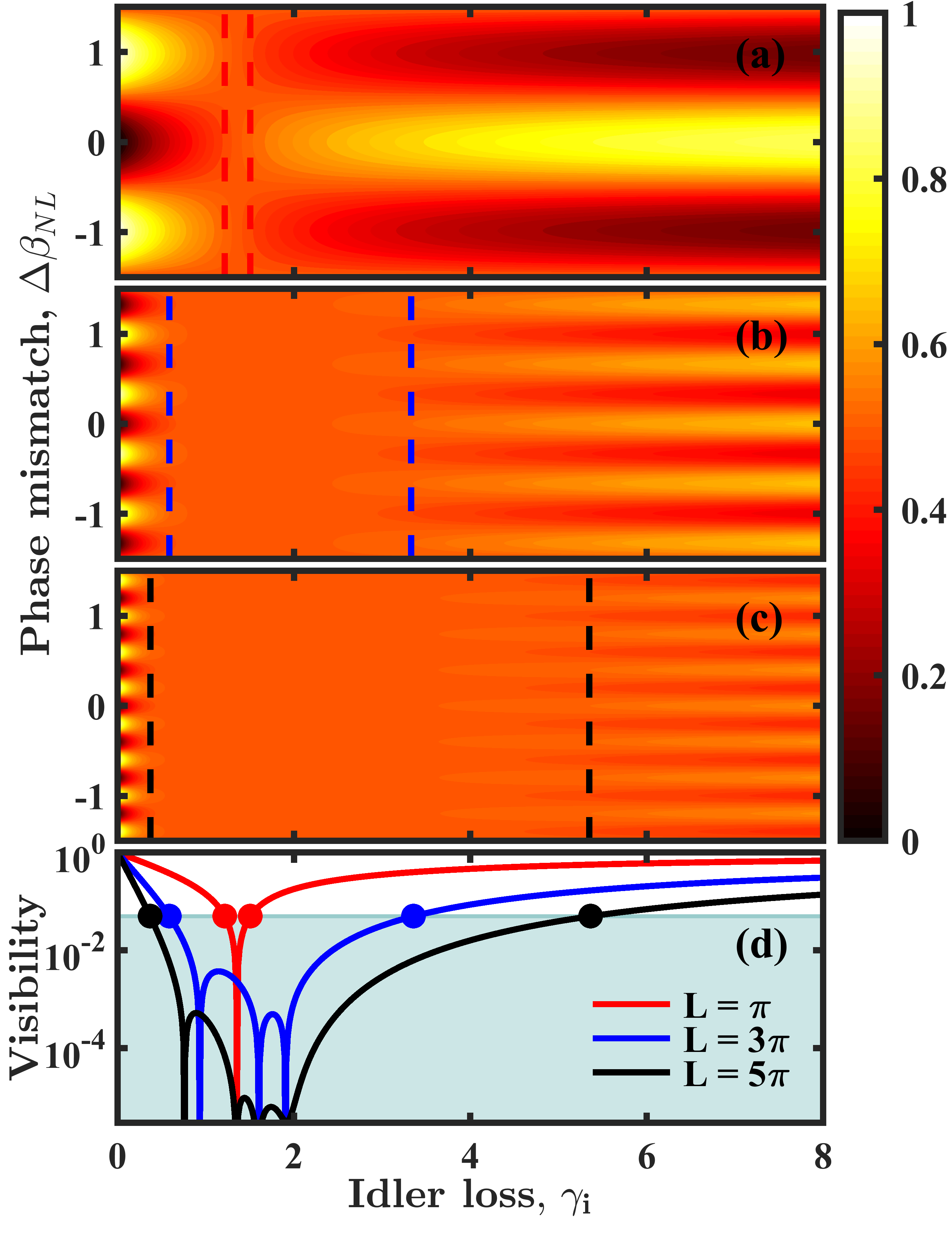}
    \caption{Signal intensity $I^s_1$ for nonlinear interferometers with PT couplers of length (a) $L = \pi$, (b) $L = 3\pi$ and (c) $L = 5\pi$. (d)~Variation of the corresponding visibility with loss $\gamma_i$. The greyed area shows region with visibility less than or equal to 0.05. 
    }
    \label{FigA2}
\end{figure}

Finally, we analyze the output signal intensity from interferometers with couplers of larger lengths to illustrate the dependence of visibility of interference fringes  on the loss $\gamma_i$. Figs.~\ref{FigA2}(a), (b) and (c) show the output $ I^s_1 $ for $ L = \pi$, $3\pi$, and $5\pi$, respectively. We plot the variation of the corresponding visibility of fringes in signal intensity in  Fig.~\ref{FigA2}(d). We have marked a threshold visibility value of 0.05 or less in this figure by the grey area to showcase the potential of the interferometer for detecting loss magnitudes in certain ranges. The loss intervals corresponding to regions between the two dashed lines in Figs.~\ref{FigA2}(a), (b) and (c) have visibility below the threshold value. We see that if $\gamma_i$ is less or more than certain threshold values (shown by dots in Fig.~\ref{FigA2}(d)), the visibility of signal interference fringes improves.     

\begin{backmatter}
\bmsection{Funding}
This work is supported by the Australian Research Council (DP190100277), the German Research Foundation (project SE 2749/1-1), the German Federal Ministry of Education and Research (project 13N14877), and the UA-DAAD  exchange  scheme (project 57559284).

\bmsection{Acknowledgment}

\bmsection{Disclosures}
The authors declare no conflicts of interest.

\bmsection{Data Availability}
Data underlying the results presented in this paper may be obtained from the authors upon reasonable request.
\end{backmatter}


\bibliography{art_PT_SPDC_spectroscopy}

\begin{thebibliography}{10}
\newcommand{\enquote}[1]{``#1''}

\bibitem{Klyshko:1988:PhotonsNonlinear}
D.~Klyshko, \href {https://doi.org/10.1201/9780203743508} {\emph{{Photons and
  Nonlinear Optics}}} (Gordon and Breach, New York, 1988).

\bibitem{Wang:1991-4614:PRA}
L.~J. Wang, X.~Y. Zou, and L.~Mandel, \enquote{Induced coherence without
  induced emission,} \href {https://doi.org/10.1103/PhysRevA.44.4614}
  {{\protect\JournalTitle{Phys. Rev. A}} \textbf{44}, 4614--4622 (1991)}.

\bibitem{Zou:1991-318:PRL}
X.~Y. Zou, L.~J. Wang, and L.~Mandel, \enquote{Induced coherence and
  indistinguishability in optical interference,} \href
  {https://doi.org/10.1103/PhysRevLett.67.318} {{\protect\JournalTitle{Phys.
  Rev. Lett.}} \textbf{67}, 318--321 (1991)}.

\bibitem{Wiseman:2000-245:PLA}
H.~M. Wiseman and K.~Molmer, \enquote{Induced coherence with and without
  induced emission,} \href {https://doi.org/10.1016/S0375-9601(00)00314-5}
  {{\protect\JournalTitle{Phys. Lett. A}} \textbf{270}, 245--248 (2000)}.

\bibitem{Lahiri:2017-33816:PRA}
M.~Lahiri, A.~Hochrainer, R.~Lapkiewicz, G.~B. Lemos, and A.~Zeilinger,
  \enquote{Partial polarization by quantum distinguishability,} \href
  {https://doi.org/10.1103/PhysRevA.95.033816} {{\protect\JournalTitle{Phys.
  Rev. A}} \textbf{95}, 033816--6 (2017)}.

\bibitem{Chekhova:2016-104:ADOP}
M.~V. Chekhova and Z.~Y. Ou, \enquote{Nonlinear interferometers in quantum
  optics,} \href {https://doi.org/10.1364/AOP.8.000104}
  {{\protect\JournalTitle{Adv. Opt. Photon.}} \textbf{8}, 104--155 (2016)}.

\bibitem{Ou:2020-80902:APLP}
Z.~Y. Ou and X.~Y. Li, \enquote{Quantum su(1,1) interferometers: Basic
  principles and applications,} \href {https://doi.org/10.1063/5.0004873}
  {{\protect\JournalTitle{APL Photonics}} \textbf{5}, 080902 (2020)}.

\bibitem{Caves:2020-1900138:ADQ}
C.~M. Caves, \enquote{Reframing su(1,1) interferometry,} \href
  {https://doi.org/10.1002/qute.201900138} {{\protect\JournalTitle{Adv. Quantum
  Technol.}} \textbf{3}, 1900138 (2020)}.

\bibitem{Ferreri:2021-461:QUA}
A.~Ferreri, M.~Santandrea, M.~Stefszky, K.~H. Luo, H.~Herrmann, C.~Silberhorn,
  and P.~R. Sharapova, \enquote{Spectrally multimode integrated su(1,1)
  interferometer,} \href {https://doi.org/10.22331/q-2021-05-27-461}
  {{\protect\JournalTitle{Quantum}} \textbf{5}, 461 (2021)}.

\bibitem{Kalashnikov:2016-98:NPHOT}
D.~A. Kalashnikov, A.~V. Paterova, S.~P. Kulik, and L.~A. Krivitsky,
  \enquote{Infrared spectroscopy with visible light,} \href
  {https://doi.org/10.1038/NPHOTON.2015.252} {{\protect\JournalTitle{Nat.
  Photon.}} \textbf{10}, 98--102 (2016)}.

\bibitem{Lemos:2014-409:NAT}
G.~B. Lemos, V.~Borish, G.~D. Cole, S.~Ramelow, R.~Lapkiewicz, and
  A.~Zeilinger, \enquote{Quantum imaging with undetected photons,} \href
  {https://doi.org/10.1038/nature13586} {{\protect\JournalTitle{Nature}}
  \textbf{512}, 409--U382 (2014)}.

\bibitem{Paterova:2018-25008:QST}
A.~V. Paterova, H.~Z. Yang, C.~W. An, D.~A. Kalashnikov, and L.~A. Krivitsky,
  \enquote{Tunable optical coherence tomography in the infrared range using
  visible photons,} \href {https://doi.org/10.1088/2058-9565/aab567}
  {{\protect\JournalTitle{Quantum Sci. Technol.}} \textbf{3}, 025008 (2018)}.

\bibitem{Paterova:2020-82:LSA}
A.~V. Paterova and L.~A. Krivitsky, \enquote{Nonlinear interference in crystal
  superlattices,} \href {https://doi.org/10.1038/s41377-020-0320-1}
  {{\protect\JournalTitle{Light Sci. Appl.}} \textbf{9}, 82 (2020)}.

\bibitem{Paterova:2017-42608:SRP}
A.~Paterova, S.~Lung, D.~A. Kalashnikov, and L.~A. Krivitsky,
  \enquote{Nonlinear infrared spectroscopy free from spectral selection,} \href
  {https://doi.org/10.1038/srep42608} {{\protect\JournalTitle{Sci. Rep.}}
  \textbf{7}, 42608 (2017)}.

\bibitem{Valles:2018-23824:PRA}
A.~Valles, G.~Jimenez, L.~J. Salazar-Serrano, and J.~P. Torres,
  \enquote{Optical sectioning in induced coherence tomography with
  frequency-entangled photons,} \href
  {https://doi.org/10.1103/PhysRevA.97.023824} {{\protect\JournalTitle{Phys.
  Rev. A}} \textbf{97}, 023824 (2018)}.

\bibitem{Lindner:2020-4426:OE}
C.~Lindner, S.~Wolf, J.~Kiessling, and F.~Kuhnemann, \enquote{{F}ourier
  transform infrared spectroscopy with visible light,} \href
  {https://doi.org/10.1364/OE.382351} {{\protect\JournalTitle{Opt. Express}}
  \textbf{28}, 4426--4432 (2020)}.

\bibitem{Kutas:2020-eaaz8065:SCA}
M.~Kutas, B.~Haase, P.~Bickert, F.~Riexinger, D.~Molter, and G.~von Freymann,
  \enquote{Terahertz quantum sensing,} \href
  {https://doi.org/10.1126/sciadv.aaz8065} {{\protect\JournalTitle{Sci. Adv.}}
  \textbf{6}, eaaz8065 (2020)}.

\bibitem{Ravaro:2008-151111:APL}
M.~Ravaro, E.~Guillotel, M.~Le~Du, C.~Manquest, X.~Marcadet, S.~Ducci,
  V.~Berger, and G.~Leo, \enquote{Nonlinear measurement of mid-infrared
  absorption in {AlO}(x) waveguides,} \href {https://doi.org/10.1063/1.2911747}
  {{\protect\JournalTitle{Appl. Phys. Lett.}} \textbf{92}, 151111 (2008)}.

\bibitem{Solntsev:2018-21301:APLP}
A.~S. Solntsev, P.~Kumar, T.~Pertsch, A.~A. Sukhorukov, and F.~Setzpfandt,
  \enquote{{{LiNbO}$_3$} waveguides for integrated {SPDC} spectroscopy,} \href
  {https://doi.org/10.1063/1.5009766} {{\protect\JournalTitle{APL Photonics}}
  \textbf{3}, 021301 (2018)}.

\bibitem{Kumar:2020-53860:PRA}
P.~Kumar, S.~Saravi, T.~Pertsch, and F.~Setzpfandt, \enquote{Integrated
  induced-coherence spectroscopy in a single nonlinear waveguide,} \href
  {https://doi.org/10.1103/PhysRevA.101.053860} {{\protect\JournalTitle{Phys.
  Rev. A}} \textbf{111}, 053860 (2020)}.

\bibitem{Ono:2019-1277:OL}
T.~Ono, G.~F. Sinclair, D.~Bonneau, M.~G. Thompson, J.~C.~F. Matthews, and
  J.~G. Rarity, \enquote{Observation of nonlinear interference on a silicon
  photonic chip,} \href {https://doi.org/10.1364/OL.44.001277}
  {{\protect\JournalTitle{Opt. Lett.}} \textbf{44}, 1277--1280 (2019)}.

\bibitem{Guo:2009-93902:PRL}
A.~Guo, G.~J. Salamo, D.~Duchesne, R.~Morandotti, M.~{Volatier-Ravat},
  V.~Aimez, G.~A. Siviloglou, and D.~N. Christodoulides, \enquote{Observation
  of {PT}-symmetry breaking in complex optical potentials,} \href
  {https://doi.org/10.1103/PhysRevLett.103.093902}
  {{\protect\JournalTitle{Phys. Rev. Lett.}} \textbf{103}, 093902 (2009)}.

\bibitem{Ruter:2010-192:NPHYS}
C.~E. Ruter, K.~G. Makris, R.~{E}l{-G}anainy, D.~N. Christodoulides, M.~Segev,
  and D.~Kip, \enquote{Observation of parity-time symmetry in optics,} \href
  {https://doi.org/10.1038/NPHYS1515} {{\protect\JournalTitle{Nat. Phys.}}
  \textbf{6}, 192--195 (2010)}.

\bibitem{El-Ganainy:2007-2632:OL}
R.~{E}l{-G}anainy, K.~G. Makris, D.~N. Christodoulides, and Z.~H. Musslimani,
  \enquote{Theory of coupled optical {PT}-symmetric structures,} \href
  {https://doi.org/10.1364/OL.32.002632} {{\protect\JournalTitle{Opt. Lett.}}
  \textbf{32}, 2632--2634 (2007)}.

\bibitem{Makris:2008-103904:PRL}
K.~G. Makris, R.~{E}l{-G}anainy, D.~N. Christodoulides, and Z.~H. Musslimani,
  \enquote{Beam dynamics in {PT} symmetric optical lattices,} \href
  {https://doi.org/10.1103/PhysRevLett.100.103904}
  {{\protect\JournalTitle{Phys. Rev. Lett.}} \textbf{100}, 103904 (2008)}.

\bibitem{El-Ganainy:2018-11:NPHYS}
R.~{E}l{-G}anainy, K.~G. Makris, M.~Khajavikhan, Z.~H. Musslimani, S.~Rotter,
  and D.~N. Christodoulides, \enquote{Non-{H}ermitian physics and {PT}
  symmetry,} \href {https://doi.org/10.1038/NPHYS4323}
  {{\protect\JournalTitle{Nat. Phys.}} \textbf{14}, 11--19 (2018)}.

\bibitem{Lin:2011-213901:PRL}
Z.~Lin, H.~Ramezani, T.~Eichelkraut, T.~Kottos, H.~Cao, and D.~N.
  Christodoulides, \enquote{Unidirectional invisibility induced by
  {PT}-symmetric periodic structures,} \href
  {https://doi.org/10.1103/PhysRevLett.106.213901}
  {{\protect\JournalTitle{Phys. Rev. Lett.}} \textbf{106}, 213901 (2011)}.

\bibitem{Hodaei:2014-975:SCI}
H.~Hodaei, M.~A. Miri, M.~Heinrich, D.~N. Christodoulides, and M.~Khajavikhan,
  \enquote{Parity-time-symmetric microring lasers,} \href
  {https://doi.org/10.1126/science.1258480} {{\protect\JournalTitle{Science}}
  \textbf{346}, 975--978 (2014)}.

\bibitem{Wiersig:2014-203901:PRL}
J.~Wiersig, \enquote{Enhancing the sensitivity of frequency and energy
  splitting detection by using exceptional points: Application to microcavity
  sensors for single-particle detection,} \href
  {https://doi.org/10.1103/PhysRevLett.112.203901}
  {{\protect\JournalTitle{Phys. Rev. Lett.}} \textbf{112}, 203901 (2014)}.

\bibitem{Hodaei:2017-187:NAT}
H.~Hodaei, A.~U. Hassan, S.~Wittek, H.~Garcia-Gracia, R.~{E}l{-G}anainy, D.~N.
  Christodoulides, and M.~Khajavikhan, \enquote{Enhanced sensitivity at
  higher-order exceptional points,} \href {https://doi.org/10.1038/nature23280}
  {{\protect\JournalTitle{Nature}} \textbf{548}, 187--200 (2017)}.

\bibitem{Ornigotti:2014-65501:JOPT}
M.~Ornigotti and A.~Szameit, \enquote{Quasi {PT}-symmetry in passive photonic
  lattices,} \href {https://doi.org/10.1088/2040-8978/16/6/065501}
  {{\protect\JournalTitle{J. Opt.}} \textbf{16}, 065501 (2014)}.

\bibitem{Main:2019-53815:PRA}
P.~B. Main, P.~J. Mosley, and A.~V. Gorbach, \enquote{Spontaneous parametric
  down-conversion in asymmetric couplers: Photon purity enhancement and
  intrinsic spectral filtering,} \href
  {https://doi.org/10.1103/PhysRevA.100.053815} {{\protect\JournalTitle{Phys.
  Rev. A}} \textbf{100}, 053815 (2019)}.

\bibitem{Su:2019-20479:OE}
J.~Su, L.~Cui, J.~M. Li, Y.~H. Liu, X.~Y. Li, and Ou, \enquote{Versatile and
  precise quantum state engineering by using nonlinear interferometers,} \href
  {https://doi.org/10.1364/OE.27.020479} {{\protect\JournalTitle{Opt. Express}}
  \textbf{27}, 20479--20492 (2019)}.

\bibitem{Li:2020-204002:APL}
J.~M. Li, J.~Su, L.~Cui, T.~Q. Xie, Z.~Y. Ou, and X.~Y. Li, \enquote{Generation
  of pure-state single photons with high heralding efficiency by using a
  three-stage nonlinear interferometer,} \href
  {https://doi.org/10.1063/5.0003601} {{\protect\JournalTitle{Appl. Phys.
  Lett.}} \textbf{116}, 204002 (2020)}.

\bibitem{Hum:2007-180:CRP}
D.~S. Hum and M.~M. Fejer, \enquote{Quasi-phasematching,} \href
  {https://doi.org/10.1016/j.crhy.2006.10.022} {{\protect\JournalTitle{C. R.
  Phys.}} \textbf{8}, 180--198 (2007)}.

\bibitem{Antonosyan:2014-43845:PRA}
D.~A. Antonosyan, A.~S. Solntsev, and A.~A. Sukhorukov, \enquote{Effect of loss
  on photon-pair generation in nonlinear waveguide arrays,} \href
  {https://doi.org/10.1103/PhysRevA.90.043845} {{\protect\JournalTitle{Phys.
  Rev. A}} \textbf{90}, 043845 (2014)}.

\bibitem{Antonosyan:2018-A6:PRJ}
D.~A. Antonosyan, A.~S. Solntsev, and A.~A. Sukhorukov, \enquote{Photon-pair
  generation in a quadratically nonlinear parity-time symmetric coupler,} \href
  {https://doi.org/10.1364/PRJ.6.0000A6} {{\protect\JournalTitle{Phot. Res.}}
  \textbf{6}, A6--A9 (2018)}.

\bibitem{Belsley:2020-28792:OE}
A.~Belsley, T.~Pertsch, and F.~Setzpfandt, \enquote{Generating path entangled
  states in waveguide systems with second-order nonlinearity,} \href
  {https://doi.org/10.1364/OE.401303} {{\protect\JournalTitle{Opt. Express}}
  \textbf{28}, 28792--28809 (2020)}.

\end{thebibliography}

\end{document}